\documentclass[english,notitlepage]{revtex4-1}
\usepackage[T1]{fontenc}
\usepackage[latin9]{inputenc}
\usepackage{xcolor}
\usepackage{pdfcolmk}
\usepackage{babel}
\usepackage{amsmath}
\usepackage{amssymb}
\usepackage{graphicx}
\PassOptionsToPackage{normalem}{ulem}
\usepackage{ulem}
\usepackage[unicode=true,
 bookmarks=false,
 breaklinks=false,pdfborder={0 0 1},backref=false,colorlinks=false]
 {hyperref}

\makeatletter

\providecolor{lyxadded}{rgb}{0,0,1}
\providecolor{lyxdeleted}{rgb}{1,0,0}

\DeclareRobustCommand{\lyxsout}[1]{\ifx\\#1\else\sout{#1}\fi}

\usepackage{babel}

\makeatother

\begin{document}
\title{Exclusive photoproduction of $B_{c}^{\pm}$ and bottomonia pairs}
\author{Sebastián Andrade, Marat Siddikov, Iván Schmidt}
\affiliation{Departamento de Física, Universidad Técnica Federico Santa María,~~~~\\
 y Centro Científico - Tecnológico de Valparaíso, Casilla 110-V, Valparaíso,
Chile}
\begin{abstract}
In this paper we analyze the photoproduction of heavy quarkonia pairs
which include $b$-quarks, such as $B_{c}^{+}B_{c}^{-}$-mesons or
charmonium-bottomonium pairs. Compared to charmonia pair production,
these channels get contributions only from some subsets of diagrams,
and thus allow for a better theoretical understanding of different production
mechanisms. In contrast to the production of hidden-flavor quarkonia,
for the production of $B_{c}$-meson pairs there are no restrictions on
internal quantum numbers in the suggested mechanisms. Using the Color
Glass Condensate approach, we estimated numerically the production
cross-sections in the kinematics of the forthcoming Electron-Proton
collider and in the kinematics of ultraperipheral collisions at LHC.
We found that the production of $J/\psi\,\eta_{c}$ and $B_{c}^{+}B_{c}^{-}$
meson pairs are the most promising channels for studies of quarkonia
pair production.
\end{abstract}
\maketitle

\section{Introduction}

Since the early days of QCD, heavy quarkonia have been used for the study
of the gluonic field in high energy interactions. Due to their heavy
masses, the heavy quarks, which constitute the quarkonia, might be
described in a perturbative approach~\cite{Korner:1991kf,Neubert:1993mb}. Nowadays
the heavy quarks processes are described in the so-called NRQCD framework, which
allows to incorporate systematically various perturbative corrections~\cite{Bodwin:1994jh,Maltoni:1997pt,Brambilla:2008zg,Feng:2015cba,Brambilla:2010cs,Cho:1995ce,Cho:1995vh,Baranov:2002cf,Baranov:2007dw,Baranov:2011ib,Baranov:2016clx,Baranov:2015laa}.
The studies of quarkonia usually focus on charmonia, since they have
significantly larger cross-sections. However, it is known that in the
charm sector there are certain technical challenges, such as for example the 
``non-universality'' of the long-distance matrix elements, which
potentially might be due to sizable corrections to the heavy quark
mass limit. In contrast, for quarkonia including bottom quarks it
is expected that the heavy quark mass approximation is much more reliable,
and thus the expected corrections should be much smaller. A special
place in these studies occupy the $B_{c}^{\pm}$-mesons, which are
made of a $b$- and $c$-quarks. As of now these states are poorly
understood, and only two states are included in the Particle Data
Group's listings. In the case of different heavy flavors, the hadronic decays
of these mesons are forbidden due to lack of phase space, which implies quite a large mean lifetime.
For the same reason, they have completely different production mechanisms
compared to the hidden-flavor states: the $B_{c}^{\pm}$ might be
produced hadronically only in hard subprocesses, which include both
$b\bar{b}$ and $\bar{c}c$ quark pairs at the partonic level. Due to
this fact, their production cross-sections are very small. Nevertheless,
studies of $B_{c}^{\pm}$ production are important for the confirmation
of our current understanding of heavy quarkonia in general, as well
as providing a potential gateway for the study of various exotic multiquark states
in their decay products.

The \emph{exclusive} production of heavy quarkonia is one of the cleanest
channels for their study. Most of the previous work on this topic focused on
single quarkonia states, due to their largest cross-section. However,
the production of multiple heavy quarkonia (e.g. heavy quarkonia pairs)
has been a subject of theoretical attention almost since inception
of QCD~\cite{Brodsky:1986ds,Lepage:1980fj,Berger:1986ii,Baek:1994kj}.
The interest in this channel has drastically increased recently due
to the forthcoming launch of high-luminosity accelerator facilities
and a recent discovery of all-heavy tetraquarks, which might be molecular
states of quarkonia pairs~\cite{Bai:2016int,Heupel:2012ua,Lloyd:2003yc,Vijande:2006vu,Vijande:2012jw,Chen:2019vrj,Esposito:2018cwh,Cardinale:2018zus,Aaij:2018zrb,Capriotti:2019huu,LHCb:2020bwg}.
Nowadays such processes might be studied both in ultraperipheral collisions
at the LHC, as well as in electron-proton collisions at the forthcoming
Electron Ion Collider (EIC)~\cite{Accardi:2012qut,DOEPR,BNLPR,AbdulKhalek:2021gbh},
the future Large Hadron electron Collider (LHeC)~\cite{AbelleiraFernandez:2012cc},
and the Future Circular Collider (FCC-he)~\cite{Mangano:2017tke,Agostini:2020fmq,Abada:2019lih}. 

Most of the previous studies of exclusive double quarkonia production
focused on the so-called two-photon mechanism, $\gamma\gamma\to M_{1}M_{2}$,
which gives the dominant contribution to the production of quarkonia
pairs with the same $C$-parity~\cite{Goncalves:2015sfy,Goncalves:2019txs,Goncalves:2006hu,Baranov:2012vu,Yang:2020xkl,Goncalves:2016ybl}.
Recently we suggested an alternative mechanism, which has significantly
larger cross-section, although leads to the production of charmonia pairs
with opposite $C$-parities. In this paper we plan to extend those
studies and analyze in detail the production of quarkonia including
$b$-quarks. For all-bottomonia pairs this study essentially repeats
our previous analysis of the charmonia sector, yet eventually gives
extremely small cross-sections. More interest present mixed charmonium-bottomonium
pairs, such as for example $J/\psi-\eta_{b}$ or $\Upsilon-\eta_{c}$,
as well as the production of $B_{c}^{+}B_{c}^{-}$ meson pairs. The production
of these states obtains contributions only from some of the diagrams
which are relevant for the production of all-charm or all-bottom quarkonia.
In the case of $B_{c}^{+}B_{c}^{-}$ production, there are no restrictions
on internal quantum numbers of the produced mesons, which presents
an important advantage over the charmonia pairs. According to theoretical
expectations, the production cross-sections in these channels are
maximal in the near-threshold region, which is relevant for searches
of different exotic states, like e.g. $b\bar{b}$-containing tetraquarks~\cite{Yang:2022cut}. 

Both in ultraperipheral collisions at LHC and in $ep$ collisions
at future colliders the dominant contribution stems from quasi-real
photons with small virtuality $Q^{2}\approx0$. At central rapidities
it is expected that the produced quarkonia will carry a just small fraction
of the colliding hadron momenta, $x_{i}\ll1$. In this kinematics
the saturation effects should play an important role in the dynamics
of partons, and thus should be properly accounted for in the theoretical
models of interaction. In what follows we'll use the the color dipole
framework, also known as Color Glass Condensate or CGC framework~\cite{GLR,McLerran:1993ni,McLerran:1993ka,McLerran:1994vd,MUQI,MV,gbw01:1,Kopeliovich:2002yv,Kopeliovich:2001ee},
which naturally incorporates the saturation effects and provides a
phenomenologically successful description of both hadron-hadron and
lepton-hadron collisions~\cite{Kovchegov:1999yj,Kovchegov:2006vj,Balitsky:2008zza,Kovchegov:2012mbw,Balitsky:2001re,Cougoulic:2019aja,Aidala:2020mzt,Ma:2014mri}.

The paper is structured as follows. Below in Section~\ref{sec:Formalism}
we present the theoretical results for the exclusive photoproduction
of heavy quarkonia pairs in the CGC approach.  In Section~\ref{sec:Numer}
we present our numerical estimates for meson pairs which include at
least one $b$ or $\bar{b}$-quark, and analyze the dependence on
quantum numbers of produced quarkonia. Finally, in Section~\ref{sec:Conclusions}
we draw conclusions.

\section{Exclusive photoproduction of meson pairs}

\label{sec:Formalism} Nowadays, photoproduction processes might
be studied both in electron-proton, proton-proton and proton-nuclear
collisions in ultraperipheral kinematics. The corresponding cross-sections
of these processes are related to photoproduction cross-section as
\begin{equation}
\frac{d\sigma_{ep\to eM_{1}M_{2}p}}{dQ^{2}\,dy_{1}d^{2}\boldsymbol{k}_{1}^{\perp}dy_{2}d^{2}\boldsymbol{k}_{2}^{\perp}}\approx\frac{\alpha_{{\rm em}}}{\pi\,Q^{2}}\,\left(1-y+\frac{y^{2}}{2}\right)\left.\frac{d\sigma_{T}\left(\gamma+p\to\gamma+p+M_{1}+M_{2}\right)}{dy_{1}d^{2}\boldsymbol{p}_{1}^{\perp}dy_{2}d^{2}\boldsymbol{p}_{2}^{\perp}}\right|_{\boldsymbol{p}_{a}^{\perp}\approx\boldsymbol{k}_{a}^{\perp}},\label{eq:LTSep-1}
\end{equation}
\begin{equation}
\frac{d\sigma_{pA\to pAM_{1}M_{2}}}{dy_{1}d^{2}\boldsymbol{k}_{1}^{\perp}dy_{2}d^{2}\boldsymbol{k}_{2}^{\perp}}=\int dn_{\gamma}\left(\omega\equiv E_{\gamma},\,\boldsymbol{q}_{\perp}\right)\,\left.\frac{d\sigma_{T}\left(\gamma+p\to\gamma+p+M_{1}+M_{2}\right)}{dy_{1}d^{2}\boldsymbol{p}_{1}^{\perp}dy_{2}d^{2}\boldsymbol{p}_{2}^{\perp}}\right|_{\boldsymbol{p}_{a}^{\perp}\approx\boldsymbol{k}_{a}^{\perp}-\boldsymbol{q}^{\perp}}\label{eq:EPA_q-2-1}
\end{equation}
where in (\ref{eq:LTSep-1}) we use standard DIS notation in which $y$ is
the inelasticity (fraction of electron energy which passes to the photon),
and $\left(y_{a},\boldsymbol{k}_{a}^{\perp}\right)$, with $a=1,2$,
are the rapidities and transverse momenta of the produced quarkonia
with respect to electron-proton or hadron-hadron collision axis. The
expression $dn_{\gamma}\left(\omega\equiv E_{\gamma},\,\boldsymbol{q}^{\perp}\right)$
in~(\ref{eq:EPA_q-2-1}) is the spectral density of the flux of equivalent
photons with energy $E_{\gamma}$ and transverse momentum $\boldsymbol{q}^{\perp}$
with respect to the nucleus, which was found explicitly in~\cite{Budnev:1975poe}.
The momenta $\boldsymbol{p}_{a}^{\perp}=\boldsymbol{k}_{a}^{\perp}-\boldsymbol{q}^{\perp}$
are the transverse parts of the quarkonia momenta with respect to
the produced photon. The nuclear form factors strongly suppress the
transverse momenta $\boldsymbol{q}^{\perp}$ larger than the inverse
nuclear radius $R_{A}^{-1}$. For this reason the average values of
$\boldsymbol{q}^{\perp}$ are quite small, $\left\langle \boldsymbol{q}_{\perp}^{2}\right\rangle \sim\left\langle Q^{2}\right\rangle \sim\left\langle R_{A}^{2}\right\rangle ^{-1}\lesssim\left(0.2\,{\rm GeV}/A^{1/3}\right)^{2}$,
and the $p_{\perp}$-dependence of the cross-sections in the left-hand
side of~(\ref{eq:EPA_q-2-1}) almost coincides with the $p_{\perp}$-dependence
of the cross-section in the integrand in the right-hand side. The
subscript letter $T$ in the right-hand side of~(\ref{eq:LTSep-1})
reminds us that the dominant contribution comes from quasireal transversely
polarized photons. The corresponding cross-section $d\sigma_{T}$
is related to the amplitude as 
\begin{align}
\frac{d\sigma_{T}}{dy_{1}\,d\left|p_{1}^{\perp}\right|^{2}dy_{2}\,d\left|p_{2}^{\perp}\right|^{2}d\phi} & \approx\frac{1}{256\pi^{4}}\left|\mathcal{A}_{\gamma_{T}p\to M_{1}M_{2}p}\right|^{2}\delta\left(\frac{p_{1}^{+}+p_{2}^{+}}{q^{+}}-1\right),\label{eq:AmpXSection-1}
\end{align}
where $\mathcal{A}_{\gamma_{T}p\to M_{1}M_{2}p}$ is the amplitude
of the exclusive process, induced by a transversely polarized photon, and
$\phi$ is the angle between the vectors $\boldsymbol{p}_{1}^{\perp}$
and $\boldsymbol{p}_{2}^{\perp}$ in transverse plane. The variable
$q^{+}$ is the light-cone momentum of the photon, and $p_{1}^{+},p_{2}^{+}$
are the light-cone momenta of the produced quarkonia.  As we will show
below, it is possible to express them via quarkonia kinematic variables
$(y_{a},\boldsymbol{p}_{a}^{\perp})$. 

For further evaluations of the amplitude $\mathcal{A}_{\gamma_{T}p\to M_{1}M_{2}p}$
it is necessary to fix the reference frame and write out explicit
light-cone momenta decompositions of the participating hadrons.
In what follows we will use the notations: $q$ for the photon momentum,
$P$ and $P'$ for the momentum of the proton before and after the
collision, and $p_{1},\,p_{2}$ for the 4-momenta of produced heavy
quarkonia. We will also use the notation $\Delta$ for the momentum
transfer to the proton, $\Delta=P'-P$, and $t$ for
its square, $t\equiv\Delta^{2}$. The light-cone expansion of the
above-mentioned momenta in the lab frame is given by
\begin{align}
q & =\,\left(q^{+},\,0,\,\,\boldsymbol{0}_{\perp}\right),\quad q^{+}\approx2E_{\gamma}\label{eq:qPhoton}\\
P & =\left(\frac{m_{N}^{2}}{2P^{-}},\,P^{-},\,\,\boldsymbol{0}_{\perp}\right),\quad P^{-}=E_{p}+\sqrt{E_{p}^{2}-m_{N}^{2}}\approx2E_{p}\\
P' & \approx\left(\frac{m_{N}^{2}+\left(\boldsymbol{p}_{1}^{\perp}+\boldsymbol{p}_{2}^{\perp}\right)^{2}}{2P^{-}-M_{1}^{\perp}e^{-y_{1}}+M_{2}^{\perp}e^{-y_{2}}},\,P^{-}-\frac{M_{1}^{\perp}e^{-y_{1}}+M_{2}^{\perp}e^{-y_{2}}}{2},\,\,-\boldsymbol{p}_{1}^{\perp}-\boldsymbol{p}_{2}^{\perp}\right),\\
p_{a} & =\left(M_{a}^{\perp}\,e^{y_{a}}\,,\,\frac{M_{a}^{\perp}e^{-y_{a}}}{2},\,\,\boldsymbol{p}_{a}^{\perp}\right),\quad a=1,2,\label{eq:MesonLC}\\
 & M_{a}^{\perp}\equiv\sqrt{M_{a}^{2}+\left(\boldsymbol{p}_{a}^{\perp}\right)^{2}},
\end{align}
where $m_{N}$ is the mass of the nucleon, and $M_{1},M_{2}$ are
the masses of produced quarkonia. In the high-energy collider kinematics,
when $q^{+},P^{-}\gg\{Q,\,M_{a},\,m_{N},\,\sqrt{|t|}\}$, there is
an approximate relation between the energy (component $q^{+}$) of
the photon and the light-cone momenta of the produced quarkonia,
\begin{align}
 & q^{+}\approx2E_{\gamma}\approx M_{1}^{\perp}\,e^{y_{1}}+M_{2}^{\perp}\,e^{y_{2}},\label{eq:qPlus}
\end{align}
which in essence reflects the fact that the change of the light-cone
plus-component proton momentum, $\left(P'\right)^{+}-P^{+}$, is negligibly
small, in agreement with the eikonal picture expectations. The relations
(\ref{eq:qPhoton}-\ref{eq:qPlus}) allow to express the quarkonia
kinematic variables $(y_{a},\boldsymbol{p}_{a}^{\perp})$ in terms of conventional
DIS variables, such as the Bjorken variable $x_{B}$ or invariant energy
$W=\sqrt{s_{\gamma p}}$. In what follows we will use these variables
$(y_{a},\boldsymbol{p}_{a}^{\perp})$, since they allow to keep explicit an
symmetry w.r.t. permutation of quarkonia, and are directly measurable
in experiments. In terms of these variables, the invariant energy $W$ of the $\gamma p$ collision
and the invariant mass $M_{12}$ of the produced heavy quarkonia pair
are given by
\begin{equation}
W^{2}\equiv s_{\gamma p}=\left(q+P\right)^{2}=-Q^{2}+m_{N}^{2}+2q\cdot P\approx-m_{N}^{2}+P^{-}\left(M_{1}^{\perp}\,e^{y_{1}}+M_{2}^{\perp}\,e^{y_{2}}\right),\label{eq:W2}
\end{equation}
and
\begin{equation}
M_{12}^{2}=\left(p_{1}+p_{2}\right)^{2}=M_{1}^{2}+M_{2}^{2}+2\left(M_{1}^{\perp}M_{2}^{\perp}\cosh\Delta y-\boldsymbol{p}_{1}^{\perp}\cdot\boldsymbol{p}_{2}^{\perp}\right)\label{eq:M12}
\end{equation}
respectively. The photoproduction amplitude $\mathcal{A}_{\gamma_{T}p\to M_{1}M_{2}p}$,
which appears in~(\ref{eq:AmpXSection-1}), is the central quantity of
interest for our study. Since the formation time of quarkonia is larger
than the typical size of the proton, the amplitude of the process
might be factorized and written as a convolution of the quarkonia
wave functions with hard amplitudes which describe photoproduction
of two quark-antiquark pairs in the gluonic field of the target. In
what follows we will refer to the heavy quarks produced in such hard
subprocess as ``final state'' quarks. The studies of exclusive production
are usually performed in the kinematics of small momenta $p_{T}$,
so we may expect that possible contributions of the color octet mechanisms~\cite{Cho:1995ce,Cho:1995vh}
are small and might be omitted. While there is no direct experimental
check of this assumption, similar studies of \emph{single} quarkonia
photoproduction in exclusive processes~\cite{Kowalski:2003hm,Kowalski:2006hc,Rezaeian:2012ji}
provide indirect evidence that this assumption might be quite reliable.

The amplitude of the double quarkonia photoproduction has been evaluated
in~\cite{Andrade:2022rbn} using the color dipole (Color Glass Condensate)
approach~\cite{GLR,McLerran:1993ka,McLerran:1994vd,MUQI,MV,gbw01:1,Kopeliovich:2002yv,Kopeliovich:2001ee}.
That evaluation was performed for the charm sector, focusing on the
production of $J/\psi\,\eta_{c}$ pairs. In this paper we are going
to extend those results for the case in which the final state quarkonia include
$b$-quarks. For the production of mixed states, such as $J/\psi\,\eta_{b},\,\text{\ensuremath{\Upsilon\,\eta_{b}}}$
and $B_{c}^{+}B_{c}^{-}$ meson pairs, only some subsets of diagrams
contribute to the total cross-section, thus providing the possibility to
understand the relative contribution of different mechanisms.

\begin{figure}
\includegraphics[scale=0.65]{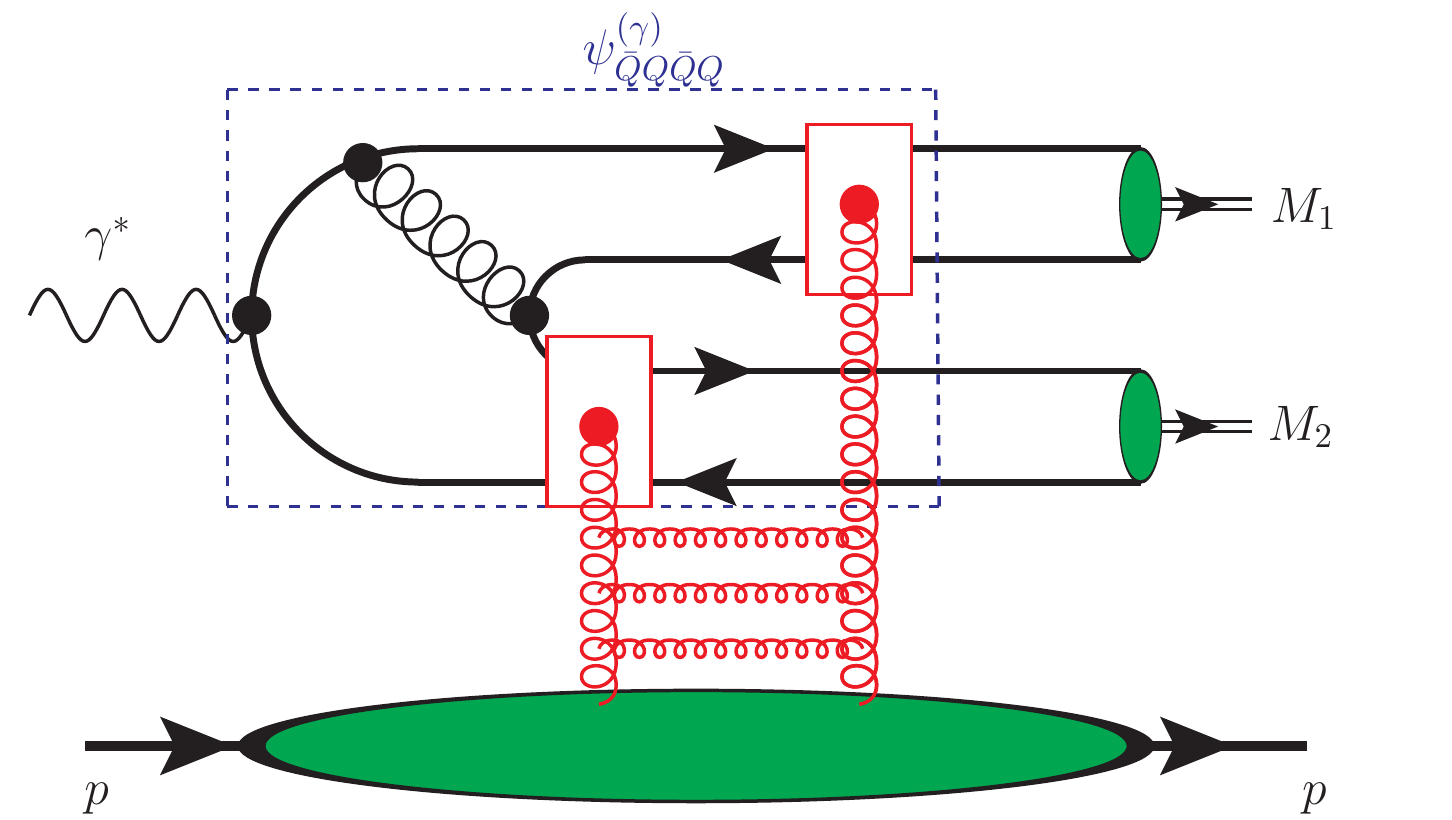}\includegraphics[scale=0.65]{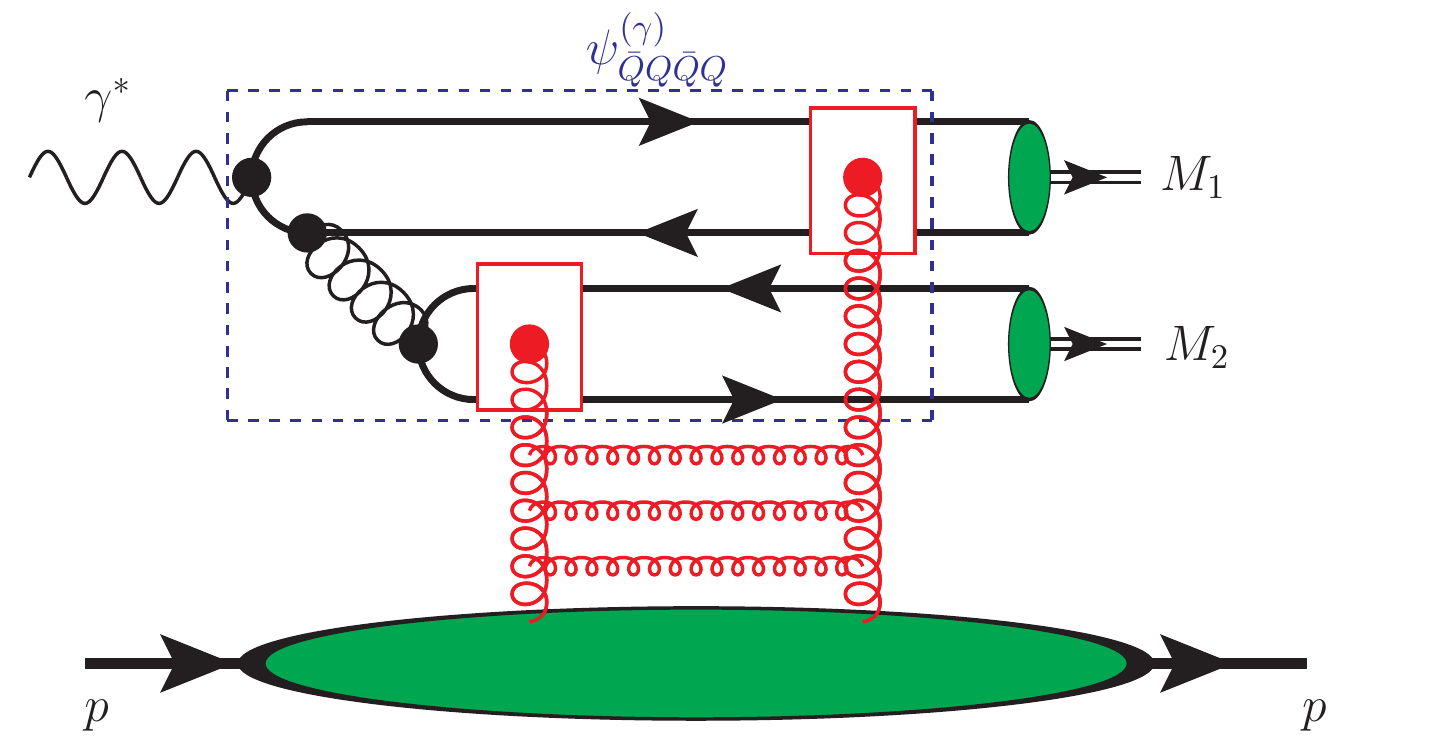}\caption{\label{fig:Photoproduction}Main classes of diagrams which contribute
in the leading order over $\alpha_{s}\left(m_{Q}\right)$ to exclusive
photoproduction of quarkonia pairs (type-$A$ and type-$B$ diagrams).
The eikonal interactions are shown schematically as exchanges of $t$-channel
gluons, indicated by the red wavy lines. In both plots it is implied:
(a) summation over all possible attachments of $t$-channel gluons
to partons inside blue dashed rectangle in upper part of diagram (b)
inclusion of diagrams with inverted direction of heavy quark lines
(\textquotedblleft charge conjugation\textquotedblright ). In the
right diagram the $t$-channel gluons must be connected to different
quark loops in order to guarantee production of \emph{color singlet}
$\bar{Q}Q$ in final states. The blue dashed rectangle schematically
shows part of the diagrams which (in absence of eikonal interactions)
would contribute to the $\bar{Q}Q\bar{Q}Q$-component of the photon
wave function $\psi_{\bar{Q}Q\bar{Q}Q}^{(\gamma)}$.}
\end{figure}

In the color dipole approach the hard process is considered in the
eikonal picture. Taking into account that the interactions of heavy
quarks with the gluonic field of the target are suppressed by the strong
coupling $\alpha_{s}\left(m_{Q}\right)$, all the leading order diagrams
might be grouped into two main classes, shown schematically in
Figure~\ref{fig:Photoproduction}. In what follows we will call them
``type-$A$'' and ``type-$B$'' respectively, and take into account
that the amplitude of the whole process might be written as an additive
sum, 
\begin{align}
\mathcal{A}\left(y_{1},\boldsymbol{p}_{T1},y_{2},\boldsymbol{p}_{T2}\right) & =\mathcal{A}^{(A)}\left(y_{1},\boldsymbol{p}_{T1},y_{2},\boldsymbol{p}_{T2}\right)+\mathcal{A}^{(B)}\left(y_{1},\boldsymbol{p}_{T1},y_{2},\boldsymbol{p}_{T2}\right),\label{eq:Amp}
\end{align}
where $\mathcal{A}^{(A)}$ and $\mathcal{A}^{(B)}$ are the contributions
of the respective classes. For the production of all-charm or all-bottom
quarkonia pairs, both $\mathcal{A}^{(A)}$ and $\mathcal{A}^{(B)}$
give nonzero contribution. In this case, $C$-parity conservation 
indicates that the produced quarkonia must have opposite $C$-parities.
For the production of mixed $B_{c}^{+}B_{c}^{-}$ meson pairs, the amplitude
$\mathcal{A}^{(B)}\equiv0$, so only the type-$A$ diagrams 
contribute. The $C$-parity conservation in this case does not impose
any constraints on the produced $B_{c}$-quarkonia internal quantum numbers,
although imposes constraints on the angular momentum $L$ of the relative motion,
which should take odd values. This constraint is relaxed if the produced
$B_{c}$ mesons have different spins, like \emph{e.g}. $B_{c}^{*+}B_{c}^{-}$
or $B_{c}^{+}B_{c}^{*-}$, where $B_{c}^{*\pm}$ is the (so far undiscovered)
vector state. Finally, for the production of charmonium-bottomonium
pairs, such as $J/\psi\,\eta_{b}$ or $\text{\ensuremath{\Upsilon\,\eta_{b}}}$,
we can see that $\mathcal{A}^{(A)}\equiv0$, so the amplitude get
contributions only of the type-$B$ diagrams. Further analysis of
the type-$B$ diagrams allows to reach some conclusions about the relative
size of mixed charmonium-bottomonium production cross-sections. Analysis
of quantum numbers suggests that a vector particle ($J^{P}=1^{-}$)
might be produced only in the upper loop (with 3 gluon attachments
to heavy quark line), whereas scalar particles might originate from
the quark loop in lower part of the diagram. This observation allows
to understand the behavior of the cross-section under permutation
of charm and bottom flavors. Since in the heavy quark mass limit each gluon attachment is suppressed
 and the natural scale for heavy quark
is its mass, we may immediately conclude that in channels with charmonium-bottomonium
production, the states with vector bottomonia are suppressed significantly
stronger than the states with vector charmonia. In the next section
we will corroborate this expectation by explicit comparison of numerical
predictions for $\text{\ensuremath{\Upsilon(1S)\,\eta_{c}}}$ and
$J/\psi\,\eta_{b}$ production cross-sections. 

In the eikonal picture the impact parameter of the parton is conserved
during interaction with the target. The interaction of the colored
dipole with the target might be described as a linear combination
of the color singlet dipole scattering amplitudes, which are known
from Deep Inelastic Scattering. For this reason, it becomes possible
to rewrite both types of diagrams as a mere convolution of the four
quark component of the photon wave function $\psi_{QQQQ}^{(\gamma)}$
with final state quarkonia wave functions and a linear combination
of color singlet dipole scattering amplitudes,

\begin{align}
\mathcal{A}^{(A)}\left(y_{1},\boldsymbol{p}_{T1},y_{2},\boldsymbol{p}_{T2}\right) & =\prod_{i=1}^{4}\left(\int d\alpha_{i}d^{2}\boldsymbol{x}_{i}\right)\delta\left(\sum_{k}\alpha_{k}-1\right)\mathcal{N}^{(A)}\left(\alpha_{1},\boldsymbol{x}_{1};\,\alpha_{2},\,\boldsymbol{x}_{2};\,\alpha_{3},\,\boldsymbol{x}_{3};\,\alpha_{4},\,\boldsymbol{x}_{4}\right)\times\label{eq:Amp-1}\\
 & \times\left[\Psi_{M_{1}}^{\dagger}\left(\alpha_{14},\,\boldsymbol{r}_{14}\right)\Psi_{M_{2}}^{\dagger}\left(\alpha_{23},\,\boldsymbol{r}_{23}\right)e^{i\left(\boldsymbol{p}_{1}^{\perp}\cdot\boldsymbol{b}_{14}+\boldsymbol{p}_{2}^{\perp}\cdot\boldsymbol{b}_{23}\right)}\delta\left(y_{1}-\mathcal{Y}_{14}\right)\delta\left(y_{2}-\mathcal{Y}_{23}\right)\right.\nonumber \\
 & +\left.\Psi_{M_{1}}^{\dagger}\left(\alpha_{23},\,\boldsymbol{r}_{23}\right)\Psi_{M_{2}}^{\dagger}\left(\alpha_{14},\,\boldsymbol{r}_{14}\right)e^{i\left(\boldsymbol{p}_{1}^{\perp}\cdot\boldsymbol{b}_{23}+\boldsymbol{p}_{2}^{\perp}\cdot\boldsymbol{b}_{14}\right)}\delta\left(y_{1}-\mathcal{Y}_{23}\right)\delta\left(y_{2}-\mathcal{Y}_{14}\right)\right]\nonumber \\
 & \times\psi_{\bar{Q}Q\bar{Q}Q}^{(\gamma)}\left(\alpha_{1},\boldsymbol{x}_{1};\,\alpha_{2},\,\boldsymbol{x}_{2};\,\alpha_{3},\,\boldsymbol{x}_{3};\,\alpha_{4},\,\boldsymbol{x}_{4};\,q\right).\nonumber 
\end{align}
\begin{align}
\mathcal{A}^{(B)}\left(y_{1},\boldsymbol{p}_{T1},y_{2},\boldsymbol{p}_{T2}\right) & =\prod_{i=1}^{4}\left(\int d\alpha_{i}d^{2}\boldsymbol{x}_{i}\right)\delta\left(\sum_{k}\alpha_{k}-1\right)\mathcal{N}^{(B)}\left(\alpha_{1},\boldsymbol{x}_{1};\,\alpha_{2},\,\boldsymbol{x}_{2};\,\alpha_{3},\,\boldsymbol{x}_{3};\,\alpha_{4},\,\boldsymbol{x}_{4}\right)\times\label{eq:Amp-2}\\
 & \times\left[\Psi_{M_{1}}^{\dagger}\left(\alpha_{12},\,\boldsymbol{r}_{12}\right)\Psi_{M_{2}}^{\dagger}\left(\alpha_{34},\,\boldsymbol{r}_{34}\right)e^{i\left(\boldsymbol{p}_{1}^{\perp}\cdot\boldsymbol{b}_{12}+\boldsymbol{p}_{2}^{\perp}\cdot\boldsymbol{b}_{34}\right)}\delta\left(y_{1}-\mathcal{Y}_{12}\right)\delta\left(y_{2}-\mathcal{Y}_{34}\right)\right.\nonumber \\
 & +\left.\Psi_{M_{1}}^{\dagger}\left(\alpha_{34},\,\boldsymbol{r}_{34}\right)\Psi_{M_{2}}^{\dagger}\left(\alpha_{12},\,\boldsymbol{r}_{12}\right)e^{i\left(\boldsymbol{p}_{1}^{\perp}\cdot\boldsymbol{b}_{34}+\boldsymbol{p}_{2}^{\perp}\cdot\boldsymbol{b}_{12}\right)}\delta\left(y_{1}-\mathcal{Y}_{34}\right)\delta\left(y_{2}-\mathcal{Y}_{12}\right)\right]\nonumber \\
 & \times\psi_{\gamma^{*}\to\bar{Q}Q\bar{Q}Q}\left(\alpha_{1},\boldsymbol{x}_{1};\,\alpha_{2},\,\boldsymbol{x}_{2};\,\alpha_{3},\,\boldsymbol{x}_{3};\,\alpha_{4},\,\boldsymbol{x}_{4};\,q\right),\nonumber 
\end{align}
where in~(\ref{eq:Amp-1},~\ref{eq:Amp-2}) $\boldsymbol{r}_{ij}=\boldsymbol{x}_{i}-\boldsymbol{x}_{j}$
is the relative distance between partons $i$ and $j$; $\alpha_{ij}=\alpha_{i}/\left(\alpha_{i}+\alpha_{j}\right)$
is the light-cone fraction carried by the quark in the pair ($ij$),
and $\boldsymbol{b}_{ij}=\left(\alpha_{i}\boldsymbol{x}_{i}+\alpha_{j}\boldsymbol{x}_{j}\right)/\left(\alpha_{i}+\alpha_{j}\right)$
is the (transverse) position of the center of mass of ($i,j$) pair.
The notations $\Psi_{M_{1}},\,\Psi_{M_{2}}$ are used for the wave
functions of the final state quarkonia $M_{1}$ and $M_{2}$ (for
a moment we disregard completely their spin indices), and $\psi_{\bar{Q}Q\bar{Q}Q}^{(\gamma)}\left(\left\{ \alpha_{i},\boldsymbol{x}_{i}\right\} ;q\right)$
is the 4-quark light-cone wave function of the virtual photon $\gamma^{*}$
which is given explicitly in Appendix~\ref{sec:WF}. The amplitudes
$\mathcal{N}^{(A)}$and $\mathcal{N}^{(B)}$ include resummation over
all possible connections of $t$-channel gluons to quark lines and,
as was shown in~\cite{Andrade:2022rbn}, can be rewritten as a linear
superposition of the color singlet dipole amplitudes $N\left(x,\boldsymbol{r}_{ij},\,\boldsymbol{b}_{ij}\right)$
\begin{align}
\mathcal{N}^{(A)} & \left(\alpha_{1},\boldsymbol{x}_{1};\,\alpha_{2},\,\boldsymbol{x}_{2};\,\alpha_{3},\,\boldsymbol{x}_{3};\,\alpha_{4},\,\boldsymbol{x}_{4}\right)=\label{eq:N_dip-1}\\
 & =\left\{ \frac{2-N_{c}^{2}}{4N_{c}}N\left(x,\,\boldsymbol{r}_{14},\,\boldsymbol{b}_{14}\right)-\frac{1}{2N_{c}}N\left(x,\,\boldsymbol{r}_{34},\,\boldsymbol{b}_{34}\right)-\frac{3+5N_{c}^{2}}{4N_{c}}N\left(x,\,\boldsymbol{r}_{12},\,\boldsymbol{b}_{12}\right)+\right.\nonumber \\
 & +\frac{1}{4N_{c}}\left[N\left(x,\,\boldsymbol{r}_{23},\,\boldsymbol{b}_{23}\right)-N\left(x,\,\frac{\alpha_{1}\boldsymbol{r}_{14}+\alpha_{3}\boldsymbol{r}_{34}}{1-\alpha_{2}},\,\boldsymbol{b}_{1344}\right)\right]\nonumber \\
 & +\frac{N_{c}^{2}+2}{4N_{c}}N\left(x,\,\boldsymbol{r}_{13},\,\boldsymbol{b}_{13}\right)+\frac{3N_{c}^{2}-2}{4N_{c}}\,N\left(x,\frac{\alpha_{1}\boldsymbol{r}_{21}+\alpha_{3}\boldsymbol{r}_{23}+\alpha_{4}\boldsymbol{r}_{24}}{1-\alpha_{2}},\boldsymbol{b}_{1234}\right)\nonumber \\
 & +\frac{3N_{c}}{2}N\left(x,\frac{\alpha_{3}\boldsymbol{r}_{13}+\alpha_{4}\boldsymbol{r}_{14}}{\alpha_{3}+\alpha_{4}},\boldsymbol{b}_{134}\right)+2N_{c}\,N\left(x,\frac{\alpha_{3}\boldsymbol{r}_{23}+\alpha_{4}\boldsymbol{r}_{24}}{\alpha_{3}+\alpha_{4}},\boldsymbol{b}_{234}\right)\nonumber \\
 & +\frac{N_{c}^{2}+1}{4N_{c}}\left[N\left(x,\frac{\alpha_{3}\boldsymbol{r}_{13}+\alpha_{4}\boldsymbol{r}_{14}}{1-\alpha_{2}},\boldsymbol{b}_{1134}\right)+N\left(x,\,\boldsymbol{r}_{24},\,\boldsymbol{b}_{24}\right)\right]\nonumber \\
 & -\frac{N_{c}}{2}\left[N\left(x,\frac{\alpha_{4}\boldsymbol{r}_{34}}{\alpha_{3}+\alpha_{4}},\boldsymbol{b}_{334}\right)+N\left(x,-\frac{\alpha_{3}\boldsymbol{r}_{34}}{\alpha_{3}+\alpha_{4}},\boldsymbol{b}_{344}\right)\right]\nonumber \\
 & -\frac{N_{c}}{2}\,N\left(x,-\frac{\alpha_{1}\left(\alpha_{3}\boldsymbol{r}_{13}+\alpha_{4}\boldsymbol{r}_{14}\right)}{\left(\alpha_{3}+\alpha_{4}\right)\left(\alpha_{1}+\alpha_{3}+\alpha_{4}\right)},\boldsymbol{b}_{34,134}\right)\nonumber \\
 & -\left.\frac{N_{c}^{2}-1}{4N_{c}}N\left(x,\,\frac{\alpha_{1}\boldsymbol{r}_{31}+\alpha_{4}\boldsymbol{r}_{34}}{1-\alpha_{2}},\,\boldsymbol{b}_{1334}\right)\right\} ,\nonumber 
\end{align}
\begin{align}
\mathcal{N}^{(B)} & \left(\alpha_{1},\boldsymbol{x}_{1};\,\alpha_{2},\,\boldsymbol{x}_{2};\,\alpha_{3},\,\boldsymbol{x}_{3};\,\alpha_{4},\,\boldsymbol{x}_{4}\right)=\label{eq:N_dip-2}\\
 & =\frac{1}{4}\left[N\left(x,\boldsymbol{r}_{23},\boldsymbol{b}_{23}\right)-N\left(x,\boldsymbol{r}_{24},\boldsymbol{b}_{24}\right)+N\left(x,\boldsymbol{r}_{3,234},\boldsymbol{b}_{2334}\right)-N\left(x,\boldsymbol{r}_{4,234},\boldsymbol{b}_{2344}\right)+\right.\nonumber \\
 & +\left.2N\left(x,\boldsymbol{r}_{14},\boldsymbol{b}_{24}\right)-2N\left(x,\boldsymbol{r}_{13},\boldsymbol{b}_{13}\right)\right]\nonumber 
\end{align}
The variables $\mathcal{Y}_{ij}$ in~(\ref{eq:Amp-1},~\ref{eq:Amp-2})
stand for the lab-frame rapidity of quark-antiquark pair made of partons
$i,\,j$. Explicitly it is given by 
\begin{equation}
\mathcal{Y}_{ij}=\ln\left(\frac{\left(\alpha_{i}+\alpha_{j}\right)q^{+}}{M_{\perp}}\right),\label{eq:Rapidity}
\end{equation}
where $\alpha_{i}$ and $\alpha_{j}$ are light-cone fractions of
the heavy quarks which form a given quarkonium.

\section{Numerical results}

\label{sec:Numer} The framework presented in the previous section
allows to make unambiguous predictions for the cross-sections. We
would like to start the presentation of numerical results with a brief discussion
of different uncertainties which are present in our evaluations. For
the sake of definiteness, we'll consider the all-charm sector and
focus on $J/\psi+\eta_{c}$ production, for which the production cross-section
is the largest (and thus is easier to study experimentally).

\begin{figure}
\includegraphics[height=8cm]{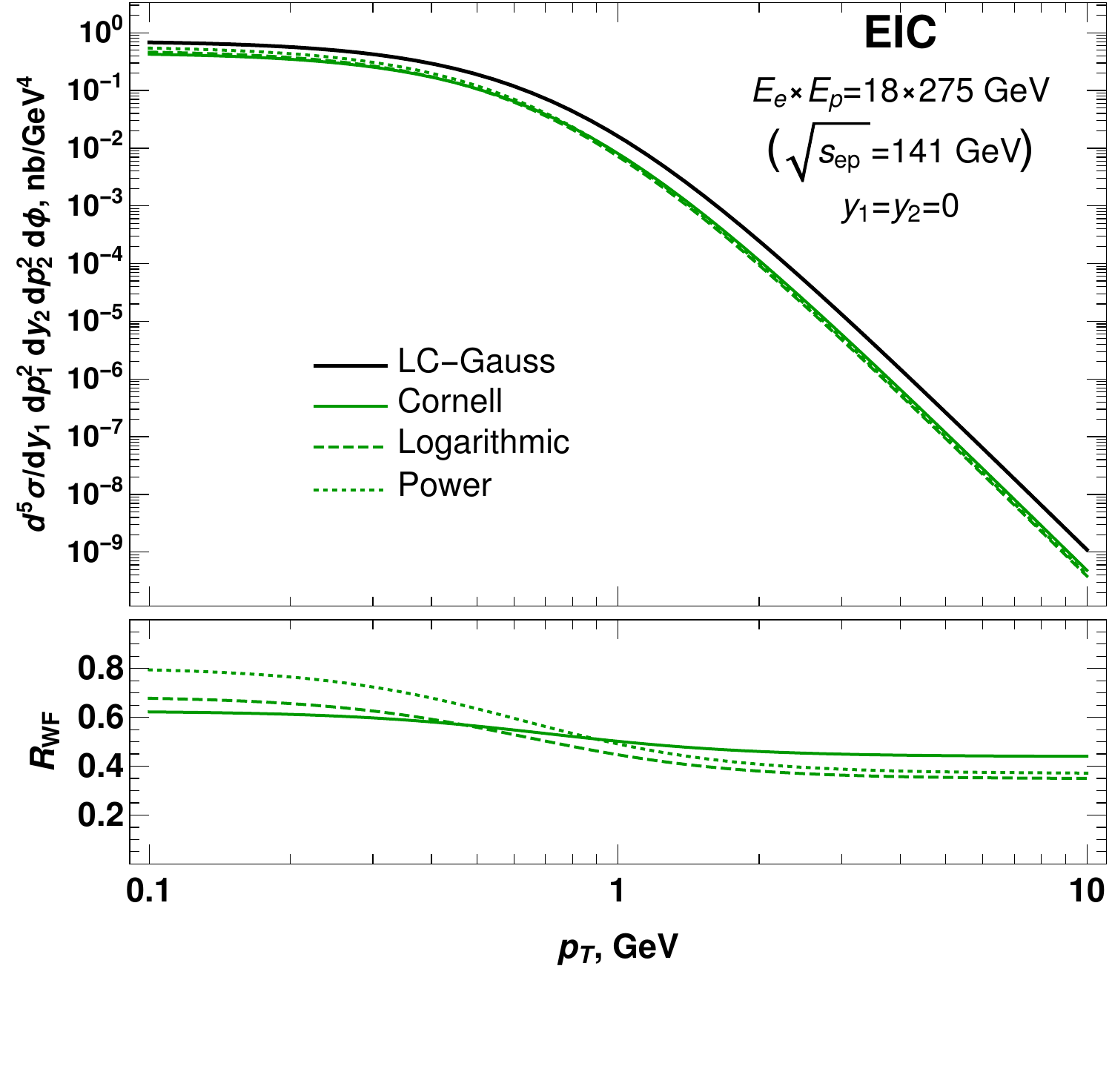}\includegraphics[height=8cm]{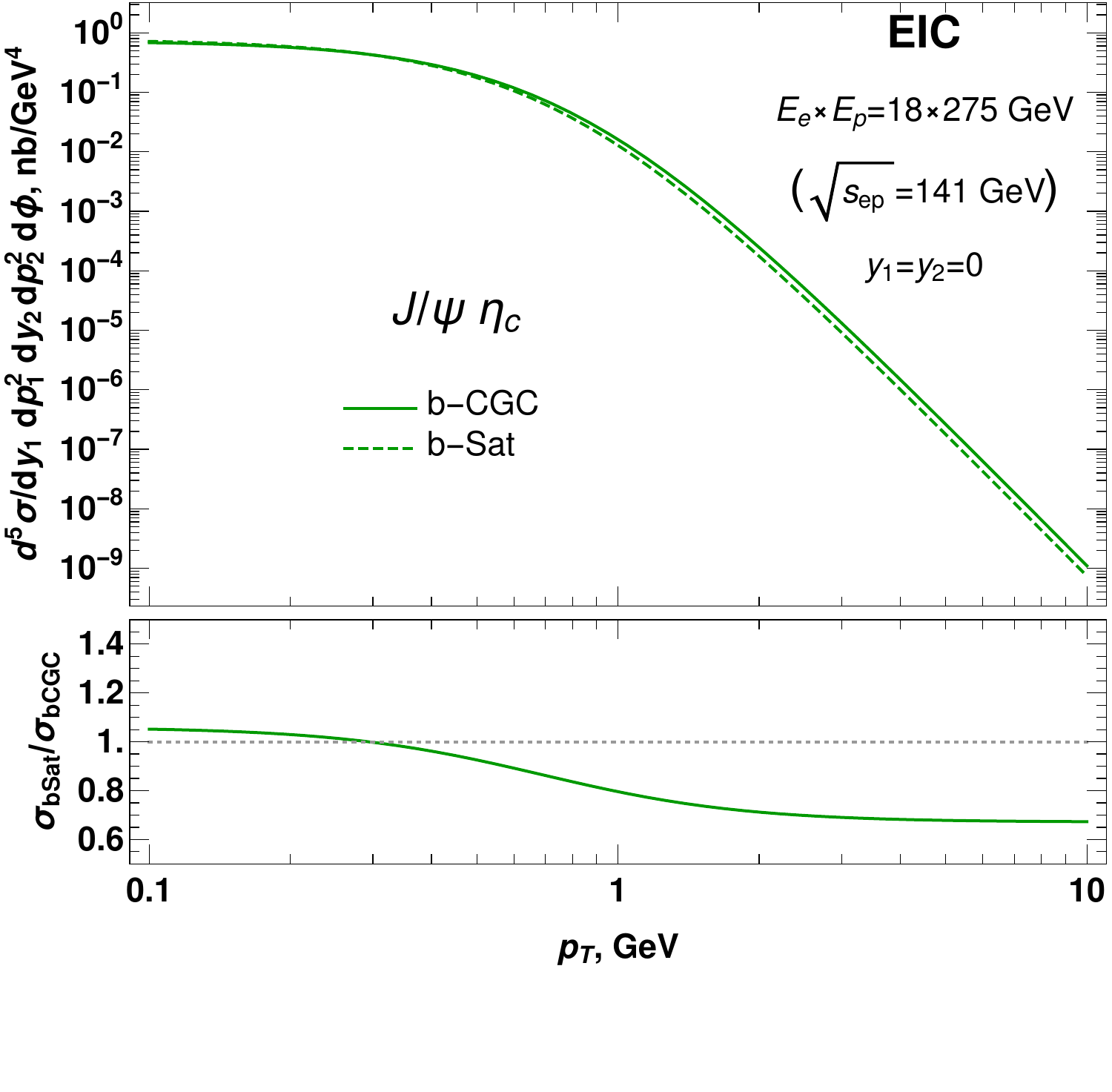}\caption{\label{fig:RelRolePhi-1}Left plot: Sensitivity of the $J/\psi\,\eta_{c}$
production cross-section to the choice of the wave function. We compare
results with the LC-Gauss parametrization of the wave function~\cite{Dosch:1996ss,BoostedGaussian}
and the wave functions evaluated in potential models~\cite{Eichten:1978tg,Eichten:1979ms,Quigg:1977dd,Martin:1980jx}.
In the lower panel of the left figure we show the ratio of the cross-sections
from the upper panel to the \textquotedblleft LC-Gauss\textquotedblright{}
curve. Right plot: Sensitivity of the $J/\psi\,\eta_{c}$ production
cross-section to the choice of to the parametrization of the dipole
cross-section. In the lower panel of the figure we show the ratio
of the cross-sections in $b$-CGC and $b$-Sat models. In both plots,
for the sake of definiteness, we considered the case when both quarkonia
are produced at central rapidities ($y_{1}=y_{2}=0$) in the lab frame;
for other rapidities and quarkonia pairs the $p_{T}$-dependence has
similar shape.}
\end{figure}
The largest source of uncertainty in our estimates is due to the wave
function of the quarkonia, which might be reformulated as uncertainty
of the Long Distance Matrix Elements (LDMEs). A popular choice used
in phenomenological estimates is the so-called light-cone Gaussian
(LC-Gauss) parametrization~\cite{Dosch:1996ss,BoostedGaussian}.
This parametrization depends on unknown parameters, which must be fixed
from phenomenology. While for $J/\psi$ and $\Upsilon$ mesons these
parameters are known or might be fixed from existing experimental
data, for heavier mesons, especially for $B_{c}$ quarkonia, this
procedure cannot be applied due to lack of experimental data, thus
making it almost impossible to make predictions for heavier mesons.
A more systematic approach is to use the wave functions of the quarkonia
evaluated in potential models, and using the well-known Brodsky-Huang-Lepage-Terentyev
(BHLT) prescription~\cite{BHT,Brodsky:2003pw,Terentev:1976jk} to
convert the rest frame wave function $\psi_{{\rm RF}}$ into a light-cone
wave function $\Psi_{{\rm LC}}$. In the small-$r$ region, which
is relevant for estimates, the wave functions of the $S$-wave heavy
quarkonia in different schemes are quite close to each other~\cite{Stadler:2018hjv,Daniel:1990ah,Kawanai:2011xb,Kawanai:2013aca},
so the uncertainty due to the choice of the potential model should
be minimal for physical observables. In order to illustrate this for
heavy quarkonia production, in the left panel of Figure~\ref{fig:RelRolePhi-1}
we compare predictions for the cross-sections obtained with the LC-Gauss
parametrization and various potential models~\cite{Eichten:1978tg,Eichten:1979ms,Quigg:1977dd,Martin:1980jx}.
The uncertainty due to the wave function does not exceed 30 per cent,
on par with expectations based on $\alpha_{s}\left(m_{c}\right)$-counting.

Another source of uncertainty in our evaluations is the choice of
parametrization of the dipole amplitude. In the right panel of the
Figure~\ref{fig:RelRolePhi-1} we compare predictions obtained with
impact parameter ($b$) dependent ``bCGC''~\cite{Kowalski:2006hc,RESH}
and ``bSat''~\cite{Rezaeian:2012ji} parametrizations of the dipole
cross-section. In the region of small $p_{T}$ both parametrizations
give very close results. In the region of very large $p_{T}$, the
difference between the two models increases due to different small-$r$
behavior implemented in ``$b$-CGC'' and ``$b$-Sat'' parametrizations:
in the former parametrization the dipole amplitude behaves like $\sim r^{\gamma}$,
whereas in the latter the dependence is much more complicated due
to built-in DGLAP evolution of gluon densities in dipole cross-section.
In what follows we will use the impact parameter ($b$) dependent
``$b$-CGC'' parametrization of the dipole cross-section~\cite{Kowalski:2006hc,RESH},

\begin{figure}
\includegraphics[width=9cm]{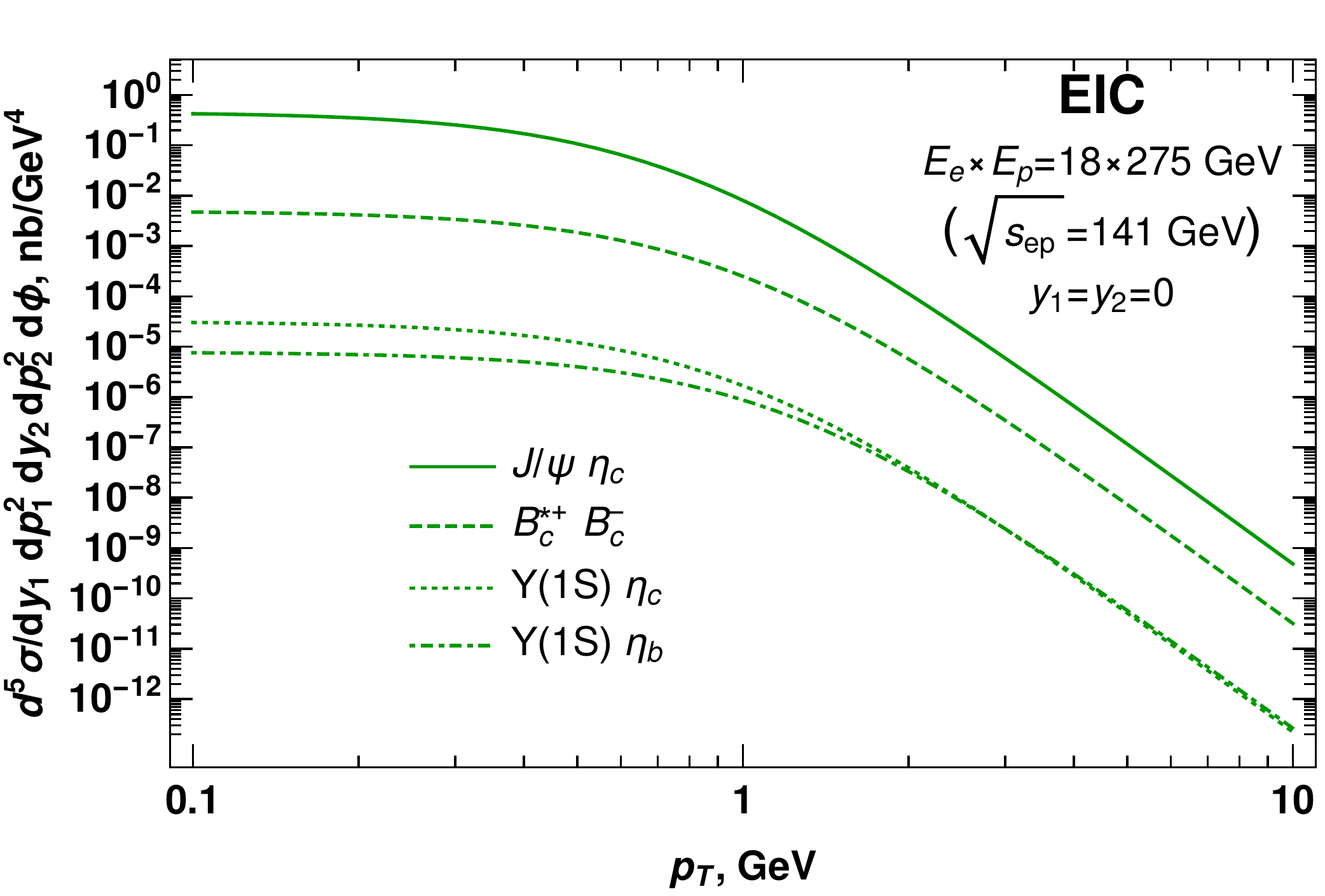}\includegraphics[width=9cm]{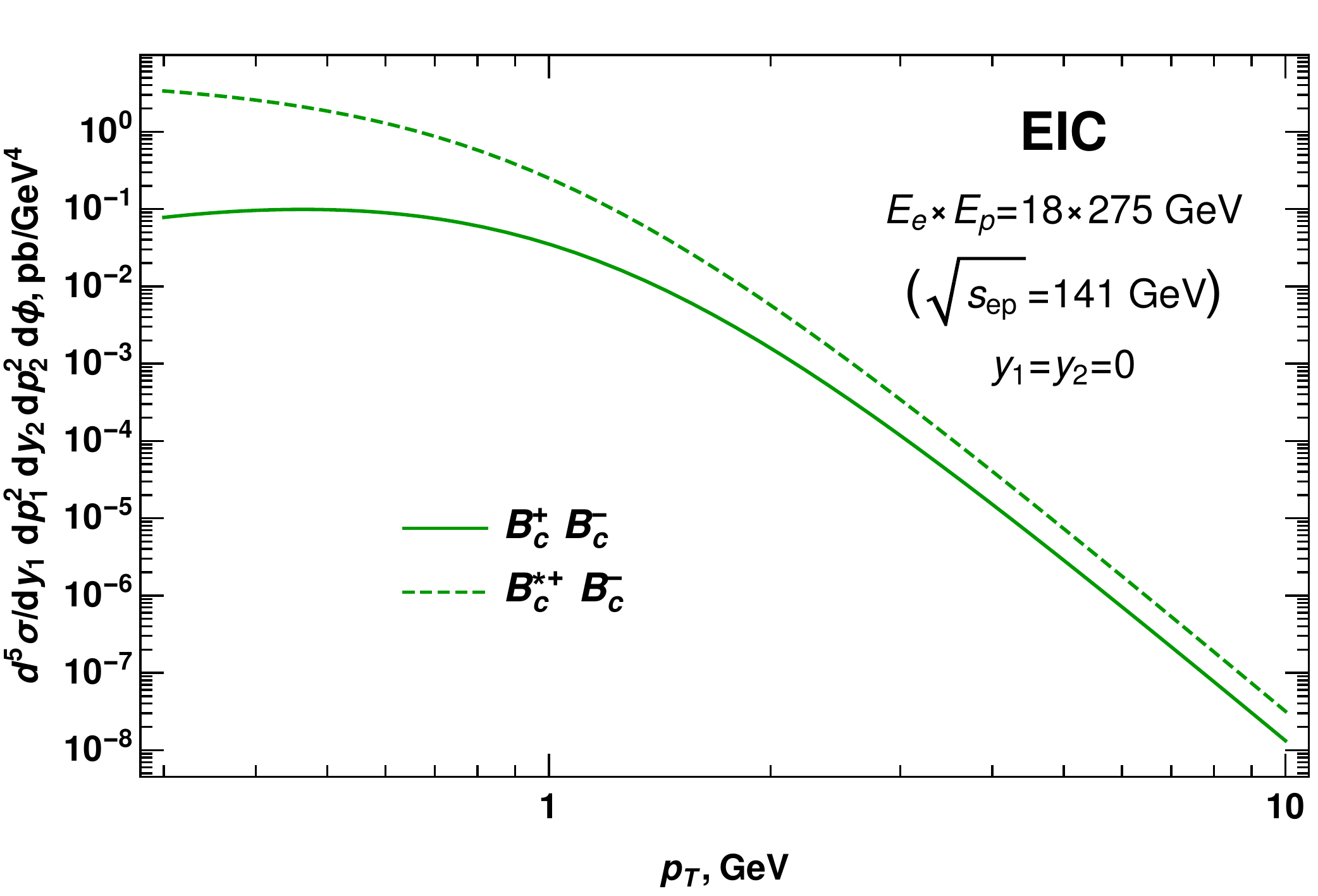}\caption{\label{fig:RelRoleJP}Left plot: Production cross-sections of different
quarkonia pairs with spin-parity $\left(1^{-},0^{-}\right)$. The
cross-sections differ significantly, due to the heavy constituents masses and
wave functions, as well as the classes of diagrams which might contribute.
Right plot: Comparison of the cross-sections for $B_{c}B_{c}$ meson
pairs with different spin-parity: $\left(1^{-},0^{-}\right)$ vs.
$\left(0^{-},0^{-}\right)$. For the sake of definiteness we considered
the case when both quarkonia are produced at central rapidities ($y_{1}=y_{2}=0$)
in the lab frame; for other rapidities the $p_{T}$-dependence has a
similar shape.}
\end{figure}

\begin{figure}
\includegraphics[height=6cm]{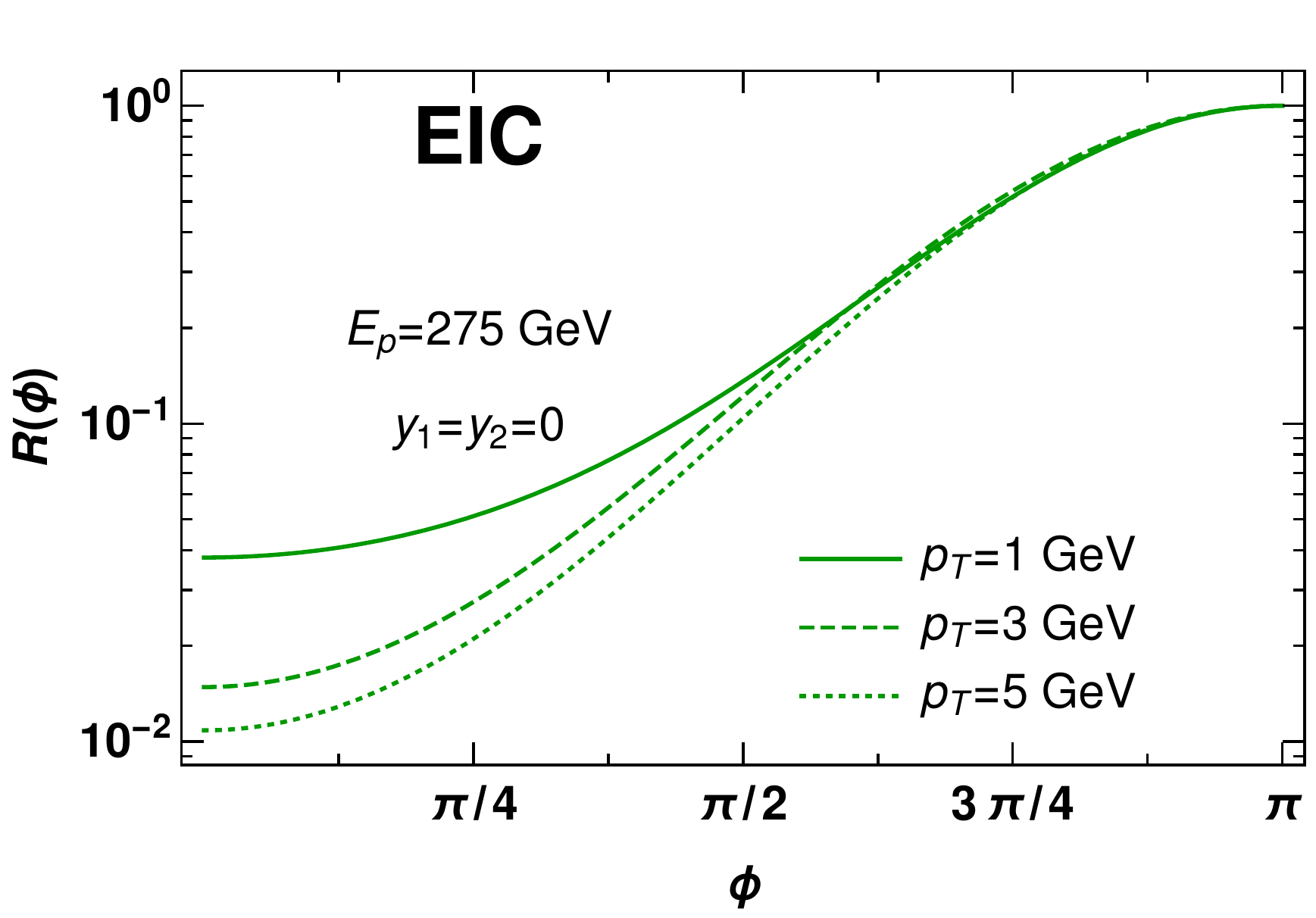}\includegraphics[height=6cm]{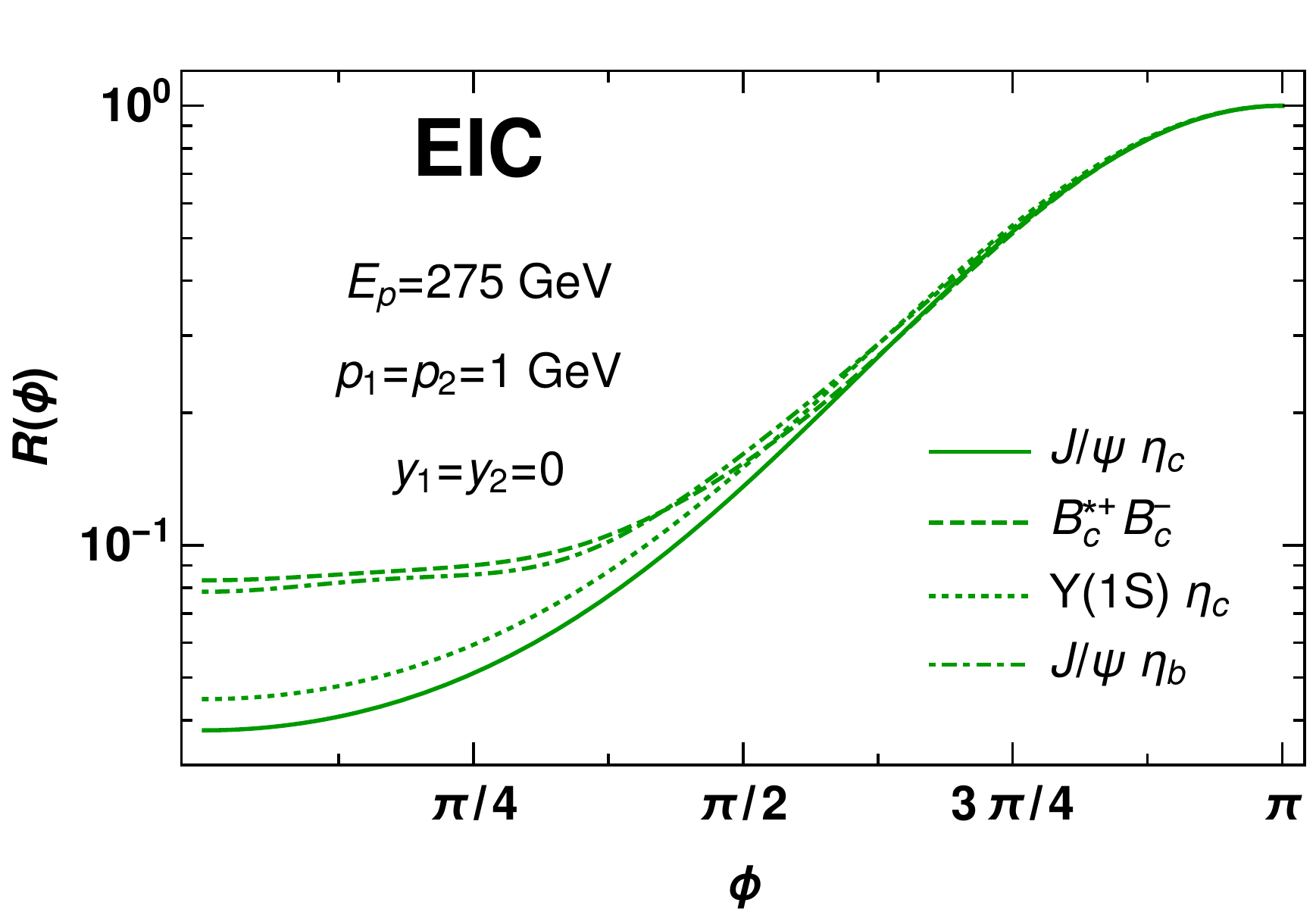}

\includegraphics[height=6cm]{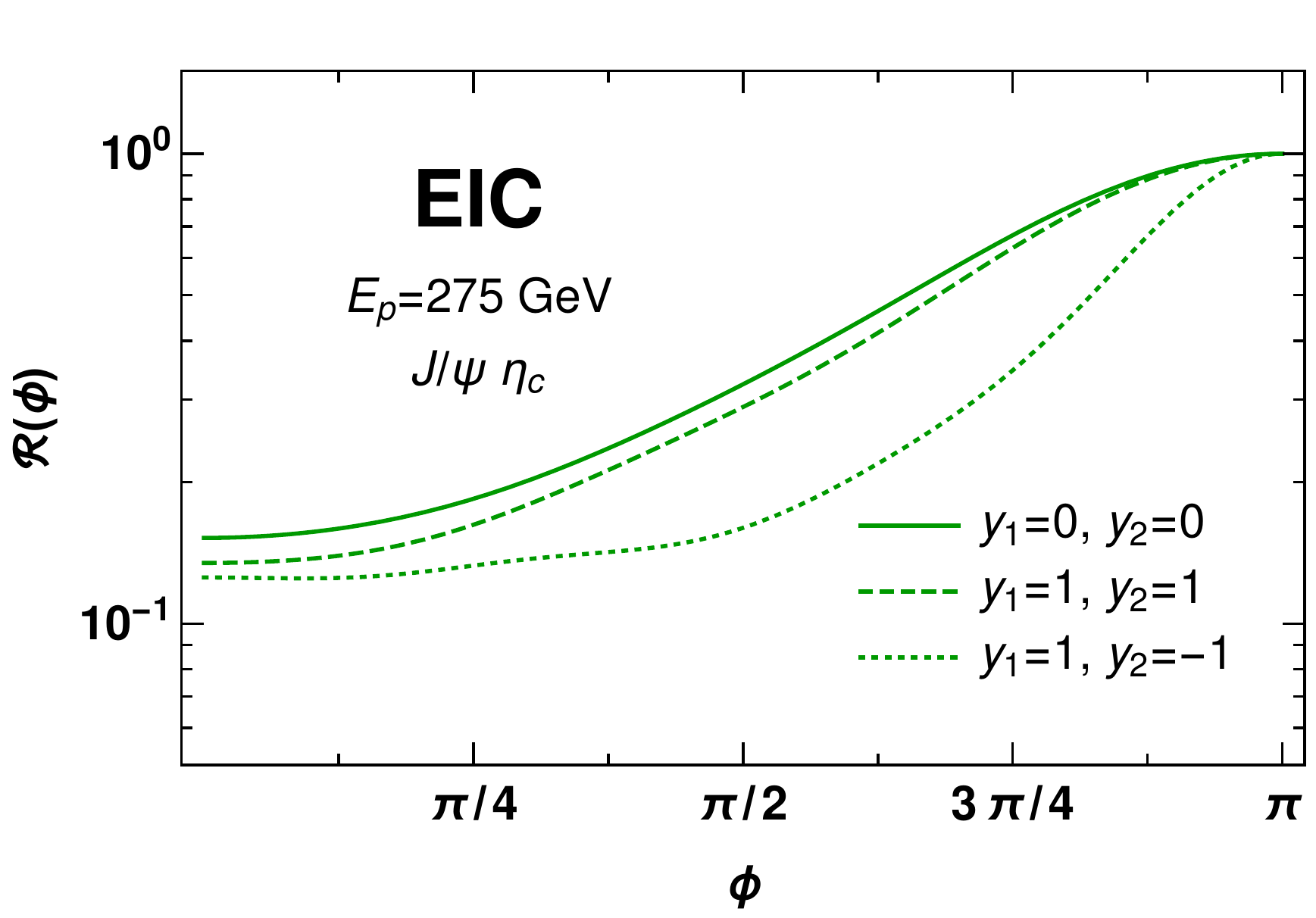}\includegraphics[height=6cm]{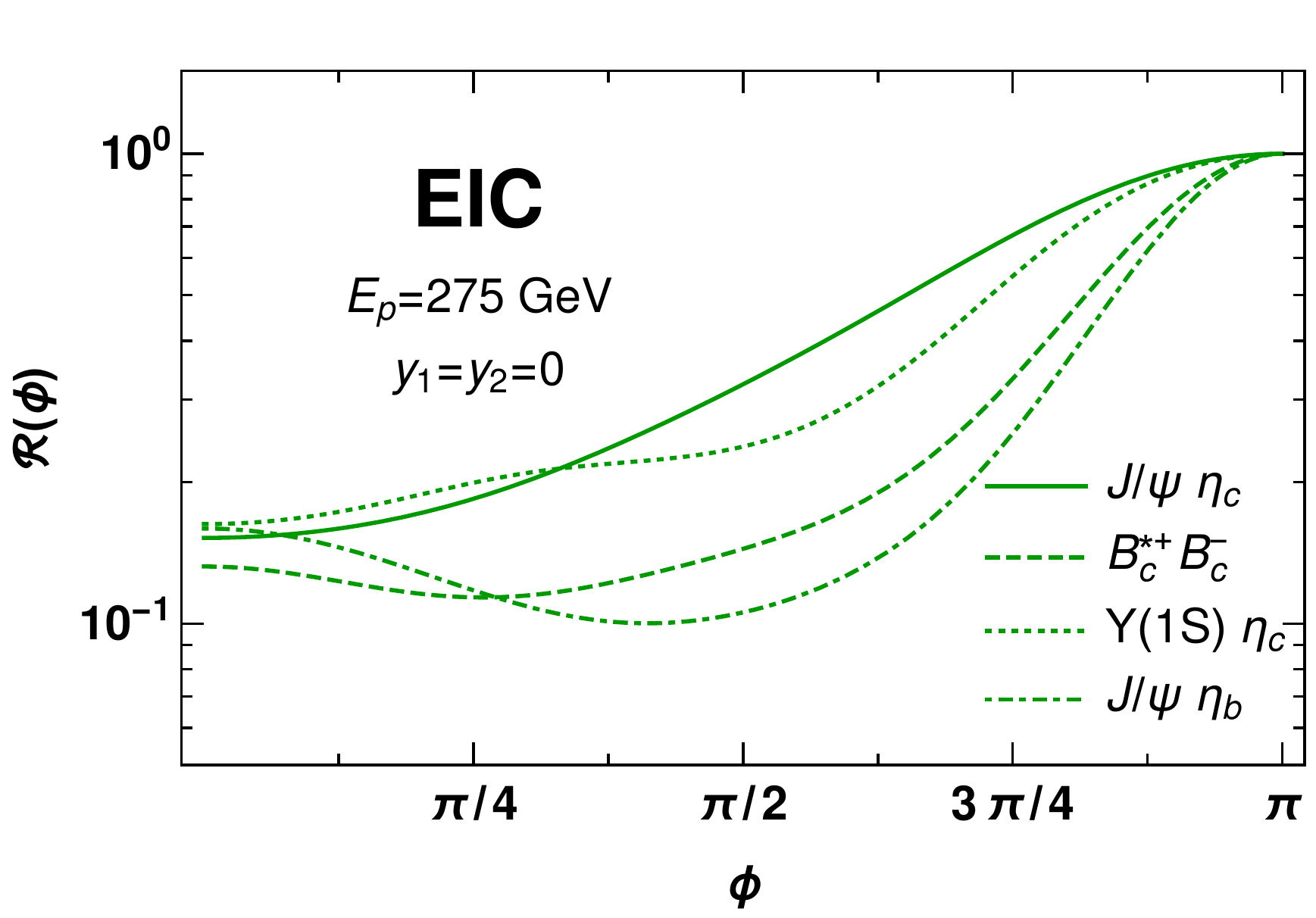}

\caption{\label{fig:RelRolePhi}Upper row: Dependence of the normalized ratio
$R(\phi)$, defined in~(\ref{eq:RatioR}), on the angle $\phi$ (difference
between azimuthal angles of both quarkonia). The left plot corresponds
to $J/\psi\,\eta_{c}$ pair production, but with different transverse
momenta. The right plot corresponds to different quarkonia states, and
fixed absolute values of the transverse momentum $p_{T}$. Lower row:
Similar dependence of $p_{T}$-integrated ratio $\mathcal{R}(\phi)$,
defined in~(\ref{eq:RatioR-1}), on the angle $\phi$. In the left
plot we compare the predictions for $J/\psi\,\eta_{c}$ with different
rapidities; in the right plot we compare the predictions for different
quarkonia pairs, at fixed central rapidity in the lab frame. The appearance
of the sharp peak in back-to-back kinematics is explained in the text.
For other rapidities the $p_{T}$-dependence has similar shape.}
\end{figure}

In Figure~\ref{fig:RelRoleJP} we illustrate the $p_{T}$-dependence
of the cross-section for different quarkonia states (for the sake
of definiteness we considered that both quarkonia are produced with
the same absolute value of transverse momenta $p_{T}$). The strong mass
dependence can be understood in the dipole picture: all gluon
interactions with dipoles of small size $\sim1/m_{Q}$ are strongly
suppressed in the heavy quark mass limit, leading to a strong suppression
of the cross-sections. As explained in the previous
section, the production of $B_{c}^{*+}B_{c}^{-}$ and charmonium-bottomonium
pairs get contributions from different classes of diagrams, which
explains the significant differences in the cross-sections. The production
of bottomonium-bottomonium pairs has significantly smaller cross-sections
and does not present any practical interest. For the $B_{c}^{+}B_{c}^{-}$
meson pairs, the $C$-parity does not impose constraints on internal
quantum numbers, and for this reason the suggested mechanism might
lead to production of both scalar and vector mesons. In the right
panel of Figure~\ref{fig:RelRolePhi} we can see that the scalar
and vector $B_{c}$ quarkonia should have similar cross-sections at
very large $p_{T}$, although might differ substantially in the region
of small momenta $p_{T}$. Potentially this channel might present
special interest for searches of the (so far undiscovered) vector
mesons $B_{c}^{*\pm}$. 

\begin{figure}
\includegraphics[height=6cm]{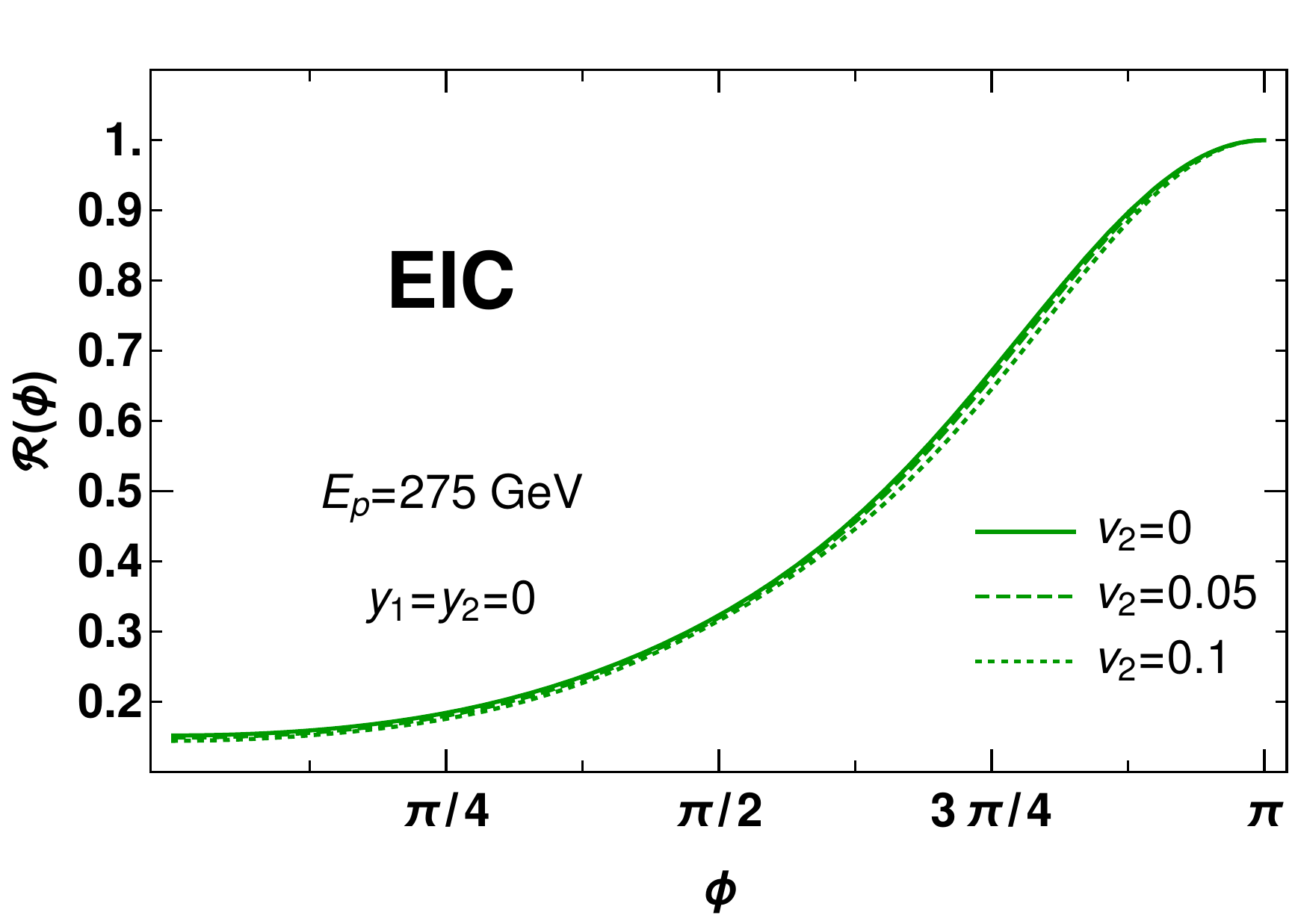}\includegraphics[height=6cm]{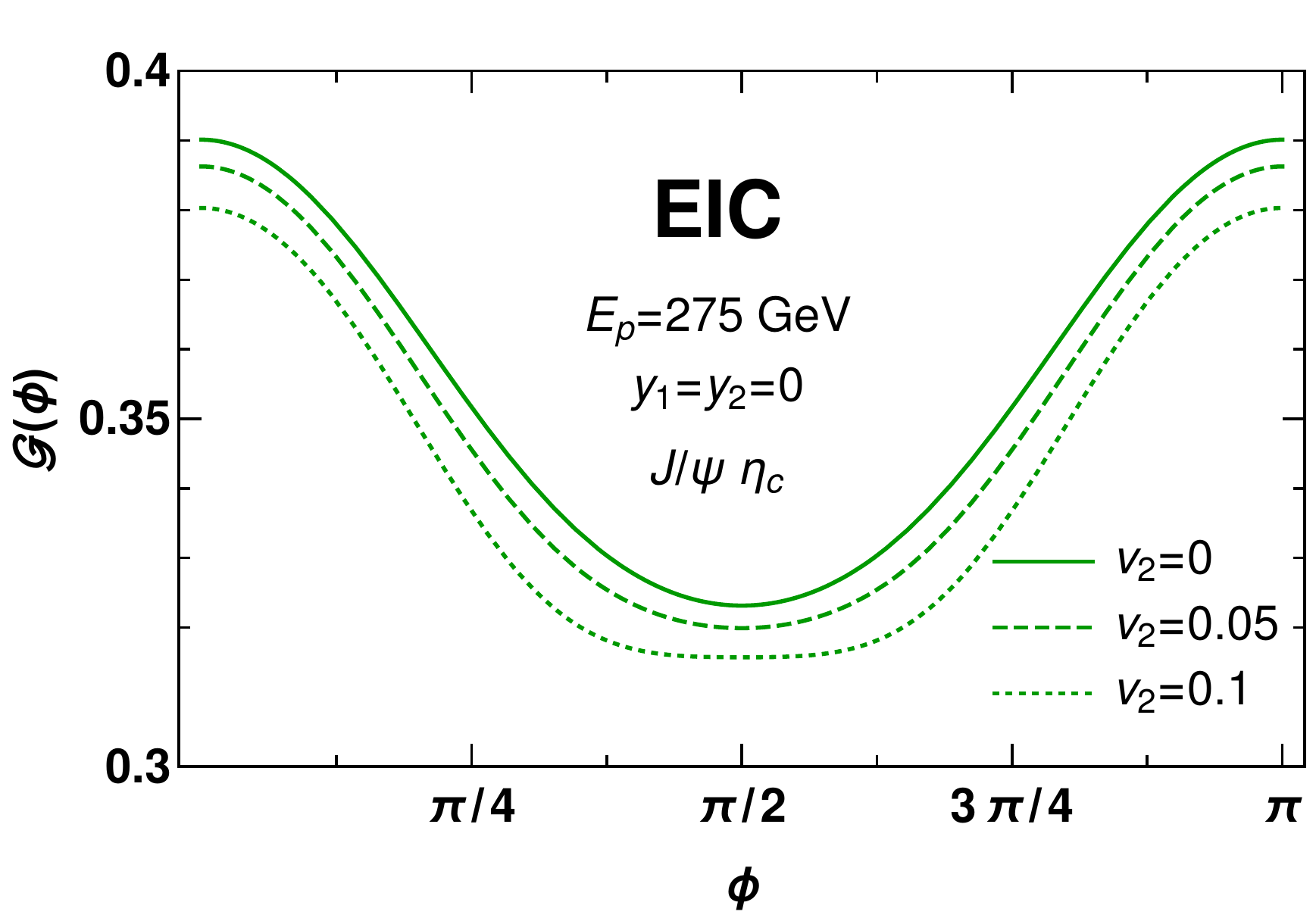}

\caption{\label{fig:RelRolePhi-2}Left plot: The dependence of the $p_{T}$-integrated
ratio $\mathcal{R}(\phi)$ defined in~(\ref{eq:RatioR-1}) on the
angle $\phi$ (azimuthal angle between quarkonia). The curves differ
by choice of the parameter $v_{2}$ defined in~(\ref{eq:NPhiDep}).
Right plot: Similar dependence for the geometric mean $\mathcal{G}(\phi)$
defined in~(\ref{eq:GeoMin}). Both plots correspond to $J/\psi\,\eta_{c}$
pair production at central rapidities ($y_{1}=y_{2}=0$). For other
quarkonia pairs the $\phi$-dependence has similar shape.}
\end{figure}

In Figure~\ref{fig:RelRolePhi} we study the dependence of the
cross-sections on the azimuthal angle $\phi$ between the transverse
momenta of $J/\psi$ and $\eta_{c}$ mesons. For the sake of definiteness,
we assumed that transverse momenta $\boldsymbol{p}_{J/\psi}^{\perp},\,\boldsymbol{p}_{\eta}^{\perp}$
of both quarkonia have equal absolute values. In order to make meaningful
comparison of the cross-sections, which differ by orders of magnitude,
in the upper row of Figure~\ref{fig:RelRolePhi} we plotted the normalized
ratio 
\begin{align}
R(\phi) & =\frac{d\sigma\left(...,\,\phi\right)/dy_{1}dp_{1}^{2}dy_{2}dp_{2}^{2}d\phi}{d\sigma\left(...,\,\phi=\pi\right)/ddy_{1}dp_{1}^{2}dy_{2}dp_{2}^{2}d\phi},\qquad R(\phi=\pi)\equiv1\label{eq:RatioR}
\end{align}
We can see that the ratio has a sharp peak in the back-to-back kinematics
($\phi=\pi$), which minimizes the momentum transfer to the target
$|t|=\left|\Delta^{2}\right|$. In contrast, for the angle $\phi\approx0$,
which maximizes the variable $|t|=\left|\Delta^{2}\right|$, the ratio
has a pronounced dip. The increase of the peak-to-trough ratio with
$p_{T}$ is due to the higher values of $|t|$ achievable in $\phi\approx0$
kinematics. For $p_{1}\not=p_{2}$ the dependence on $\phi$ is qualitatively
similar, although the maximum and minimum are less pronounced. The dependence
on $\phi$ has very similar shape for all quarkonia states. Due to
smallness of the cross-sections at large $p_{T}$, it could be challenging
to measure the ratio~(\ref{eq:RatioR}). For this reason, we also
analyzed the ratio of the $p_{T}$-integrated cross-sections 
\begin{align}
\mathcal{R}(\phi) & =\frac{d\sigma\left(...,\,\phi\right)/dy_{1}dy_{2}d\phi}{d\sigma\left(...,\,\phi=\pi\right)/ddy_{1}dy_{2}d\phi},\qquad\mathcal{R}(\phi=\pi)\equiv1\label{eq:RatioR-1}
\end{align}
which should be easier to study experimentally. Its $\phi$-dependence
is qualitatively similar to that of~(\ref{eq:RatioR}), although is
milder. This happens because the $p_{T}$-integrated cross-sections
get a dominant contribution from the region of relatively small $p_{T}$,
for which the momentum transfer $t$ remains small for all angles
$\phi$. We expect that experimental study of the ratios~(\ref{eq:RatioR},\ref{eq:RatioR-1})
could help to understand possible correlations between orientations
of the dipole separation vector $\boldsymbol{r}$ and dipole impact
parameter $\boldsymbol{b}$ in a color singlet dipole amplitude $N\left(x,\,\boldsymbol{r},\,\boldsymbol{b}\right)$.
Such dependence is frequently neglected in phenomenological parametrizations,
like $b$-CGC and $b$-Sat, and for many channels (\emph{e.g}. DIS,
DVCS, DVMP) this simplification is justified, since the corresponding
cross-sections are not sensitive to the $\phi$-dependence. However,
in different theoretical models it has been demonstarted that such
dependence might exist, and its extraction from data becomes possible
if the final state includes \emph{two} hadrons in addition to recoil
proton (see~\cite{Mantysaari:2019csc,Hatta:2020bgy} for more details).
While all previous studies of this dependence focused on exclusive
dijet production, the exclusive production of heavy quarkonia pairs
might be also used for this purpose and presents a lot of interest
in view of its very clean final state. In order to illustrate feasibility
of such measurement, we analyze the modification of the $\phi$-dependence
of the ratios~(\ref{eq:RatioR},\ref{eq:RatioR-1}) due to possible
angular dependence of the dipole amplitude. Following~\cite{Mantysaari:2019csc},
temporarily we'll assume that such dependence is given by the dipole
amplitude
\begin{equation}
N\left(x,\boldsymbol{r},\boldsymbol{b}\right)\approx N_{b{\rm -CGC}}\left(x,\,r,\,b\right)\left(1+2v_{2}\cos\left(2\theta_{\boldsymbol{r},\boldsymbol{b}}\right)\right)\label{eq:NPhiDep}
\end{equation}
where $\theta_{\boldsymbol{r},\boldsymbol{b}}$ is the angle between
vectors $\boldsymbol{r}$ and $\boldsymbol{b}$, and $v_{2}$ is a
numerical constant which characterizes the size of angular dependent
term. In the left panel of the Figure~\ref{fig:RelRolePhi-2} we
illustrate the $\phi$-dependence of the ratio $\mathcal{R}(\phi)$
for different values of $v_{2}$. Since expected values of $v_{2}$
are very small (of order a few percent), the shape of $\mathcal{R}(\phi)$
experiences only small changes. For this reason, extraction of the
constant $v_{2}$ from quarkonia pair production requires to use special
observables which would enhance sensitivity to $v_{2}$. We suggest
to use for this purpose the geometric mean
\begin{equation}
\mathcal{G}(\phi)=\sqrt{\mathcal{R}(\phi)\mathcal{R}(\pi-\phi)}.\label{eq:GeoMin}
\end{equation}
The strong $\phi$-dependence of the cross-sections, which is due
to increase of momentum transfer $t$ to the recoil proton, largely
cancels in $\mathcal{G}(\phi)$, and thus
extraction of $v_{2}$ from this observable might be done wih better
precision. Indeed, for small $t$, the dependence of the cross-sections
on $t$ might be approximated with exponent,
\begin{equation}
R(\phi)\sim e^{B\,t}\sim e^{-B\left(\boldsymbol{p}_{1}^{\perp}+\boldsymbol{p}_{2}^{\perp}\right)^{2}}\sim e^{-B\left[\left(p_{1}^{\perp}\right)^{2}+\left(p_{1}^{\perp}\right)^{2}\right]}e^{-2Bp_{1}^{\perp}p_{2}^{\perp}\cos\phi}.
\end{equation}
In the product $R(\phi)R(\pi-\phi)$ the exponents with $\phi$-dependence
cancel, thus giving possibility to study the ``residual'' $\phi$-dependence
due to $\mathcal{O}\left(v_{2}\right)$-terms in prefactors. The extension
of this proof for the $p_{T}$-integrated ratios, which appear in~(\ref{eq:GeoMin}),
is straightforward. As we can see from the right panel of the Figure~\ref{fig:RelRolePhi-2},
the observable $\mathcal{G}(\phi)$ indeed has significantly milder dependence on $\phi$, and thus is much better suited for extraction of $v_{2}$.

\begin{figure}
\includegraphics[width=9cm]{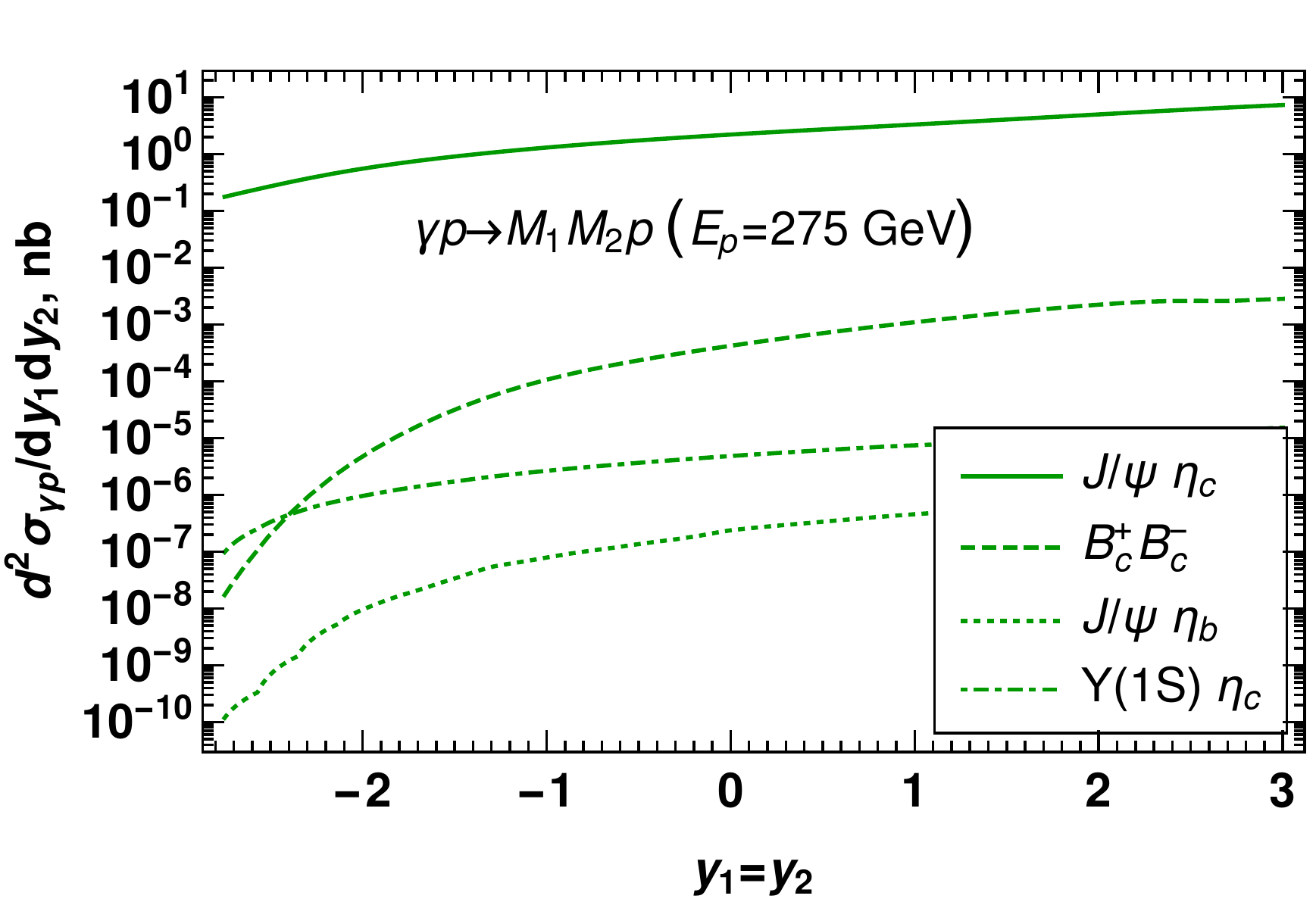}\includegraphics[width=9cm]{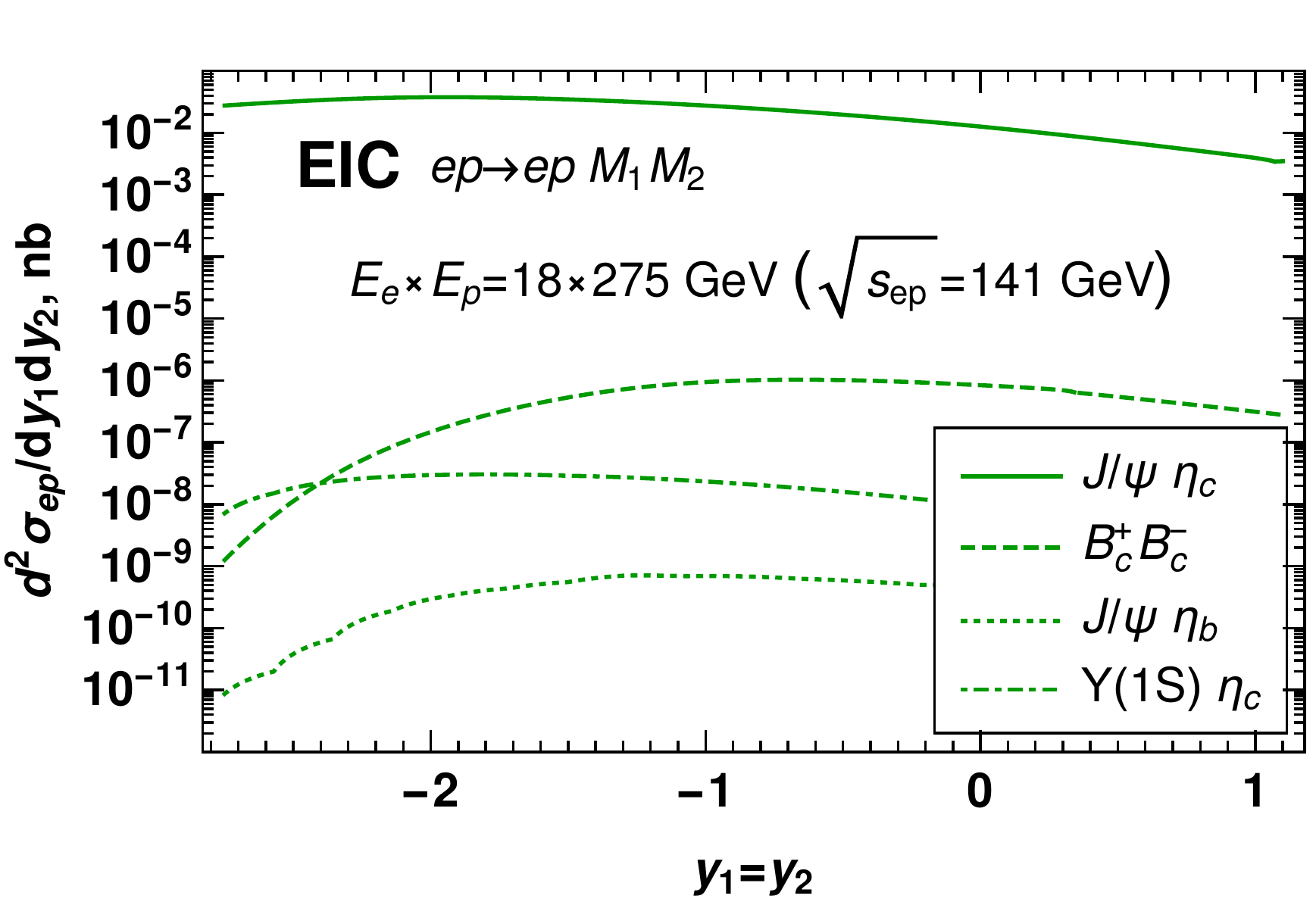}

\includegraphics[width=9cm]{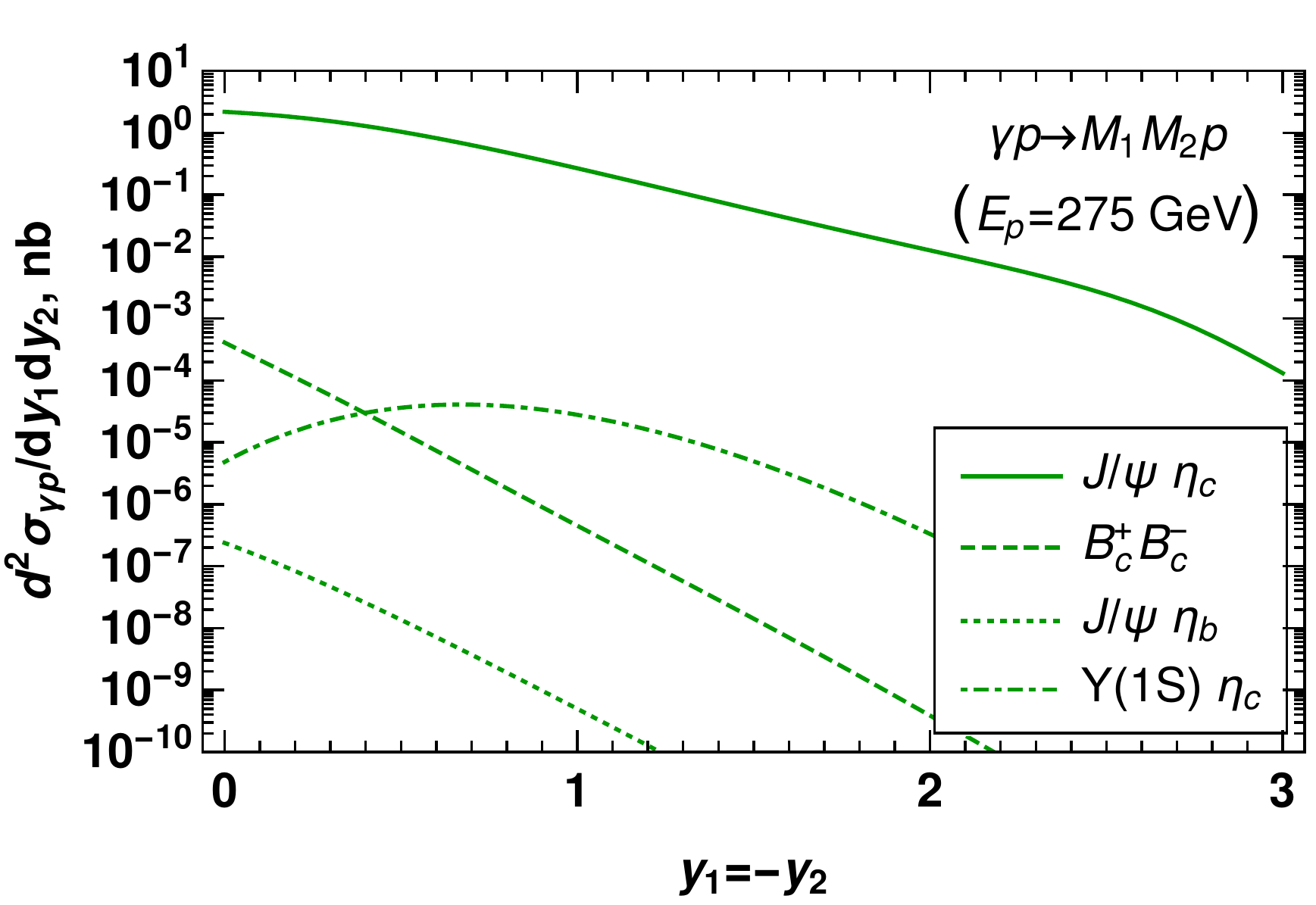}\includegraphics[width=9cm]{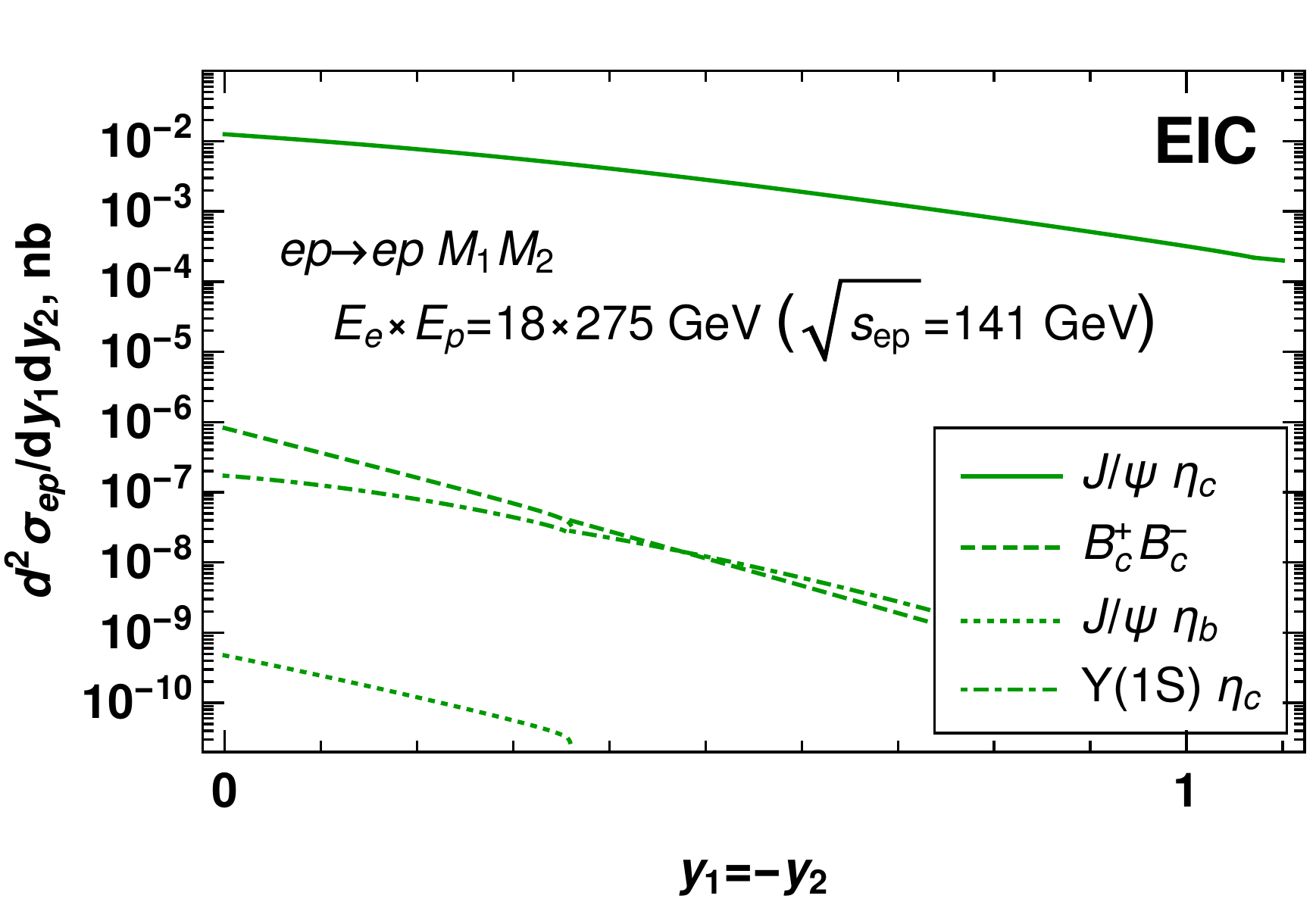}

\caption{\label{fig:Rapidities} The rapidity dependence of the photoproduction
cross-section in EIC kinematics. The plots in the upper row correspond
to a configuration with equal rapidities of the produced quarkonia,
$y_{1}=y_{2}$, whereas the lower row corresponds to rapidities which
differ by a sign in the lab frame, $y_{1}=-y_{2}$. In both rows the left
column corresponds to cross-section of photoproduction subprocess,
$\gamma p\to M_{1}M_{2}p$, whereas the right column corresponds to
a cross-section of the electroproduction process $ep\to M_{1}M_{2}ep$
and takes into account an additional leptonic factor, as defined in~(\ref{eq:LTSep-1}).}
\end{figure}

\begin{figure}
\includegraphics[width=9cm]{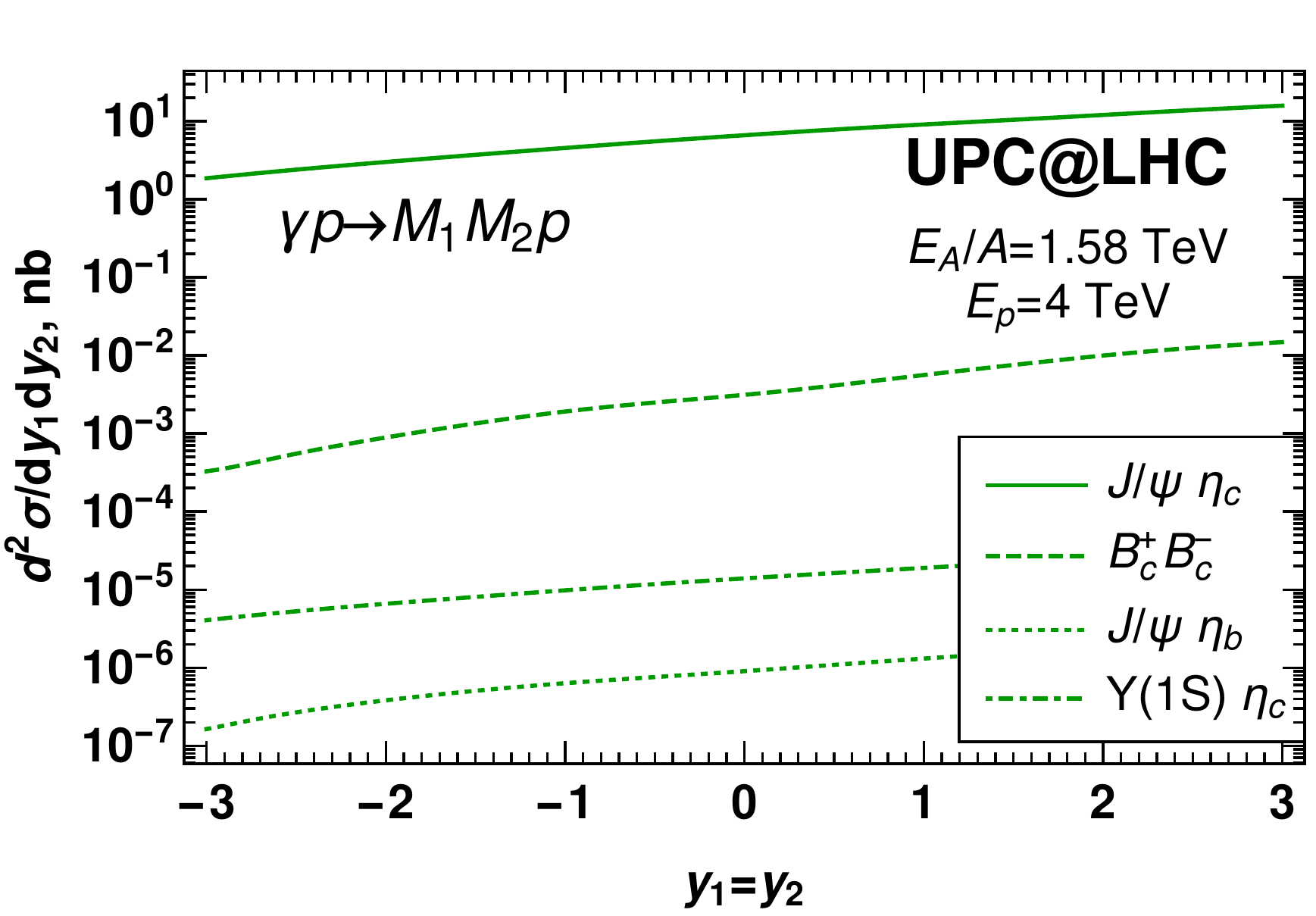}\includegraphics[width=9cm]{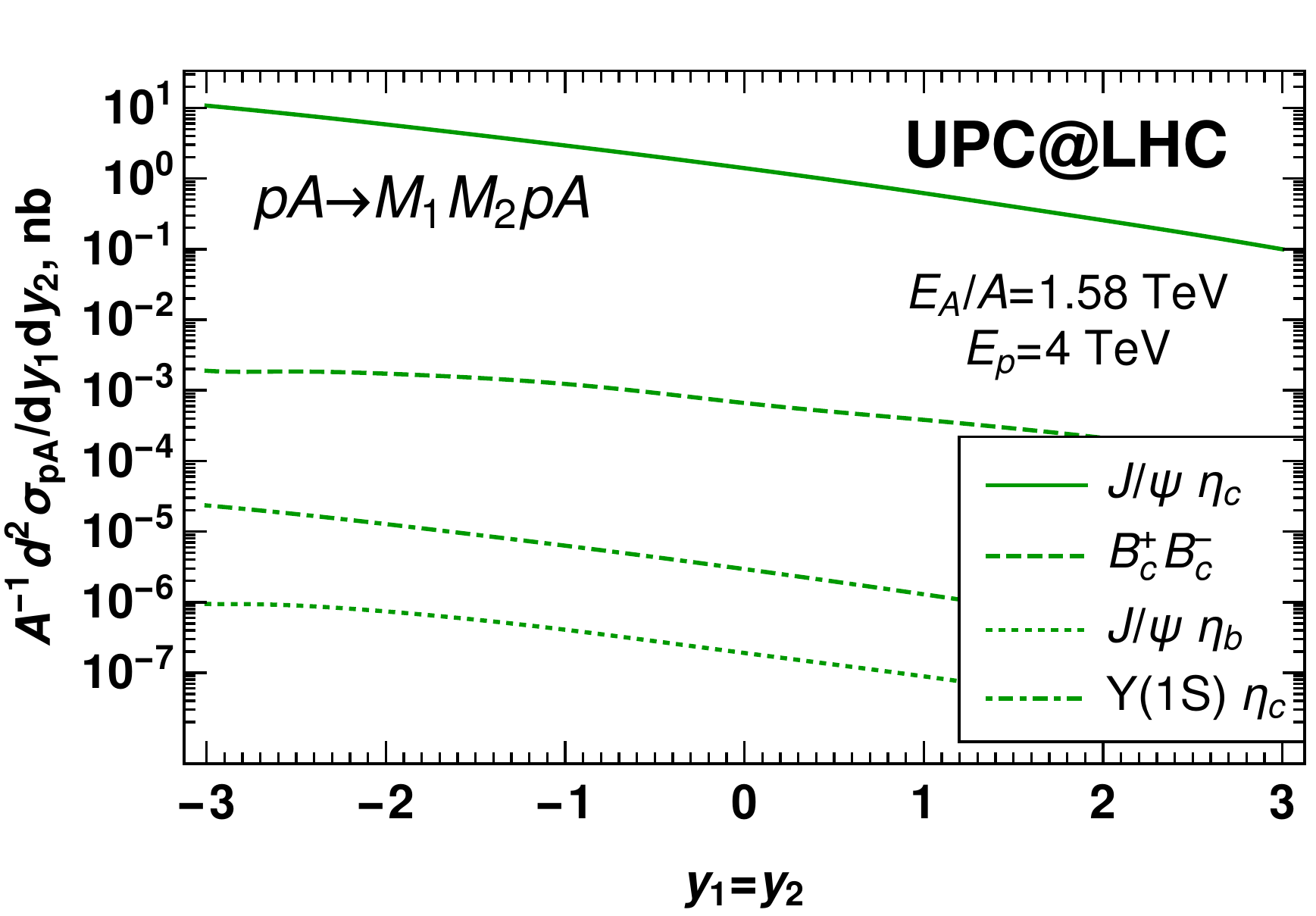}

\includegraphics[width=9cm]{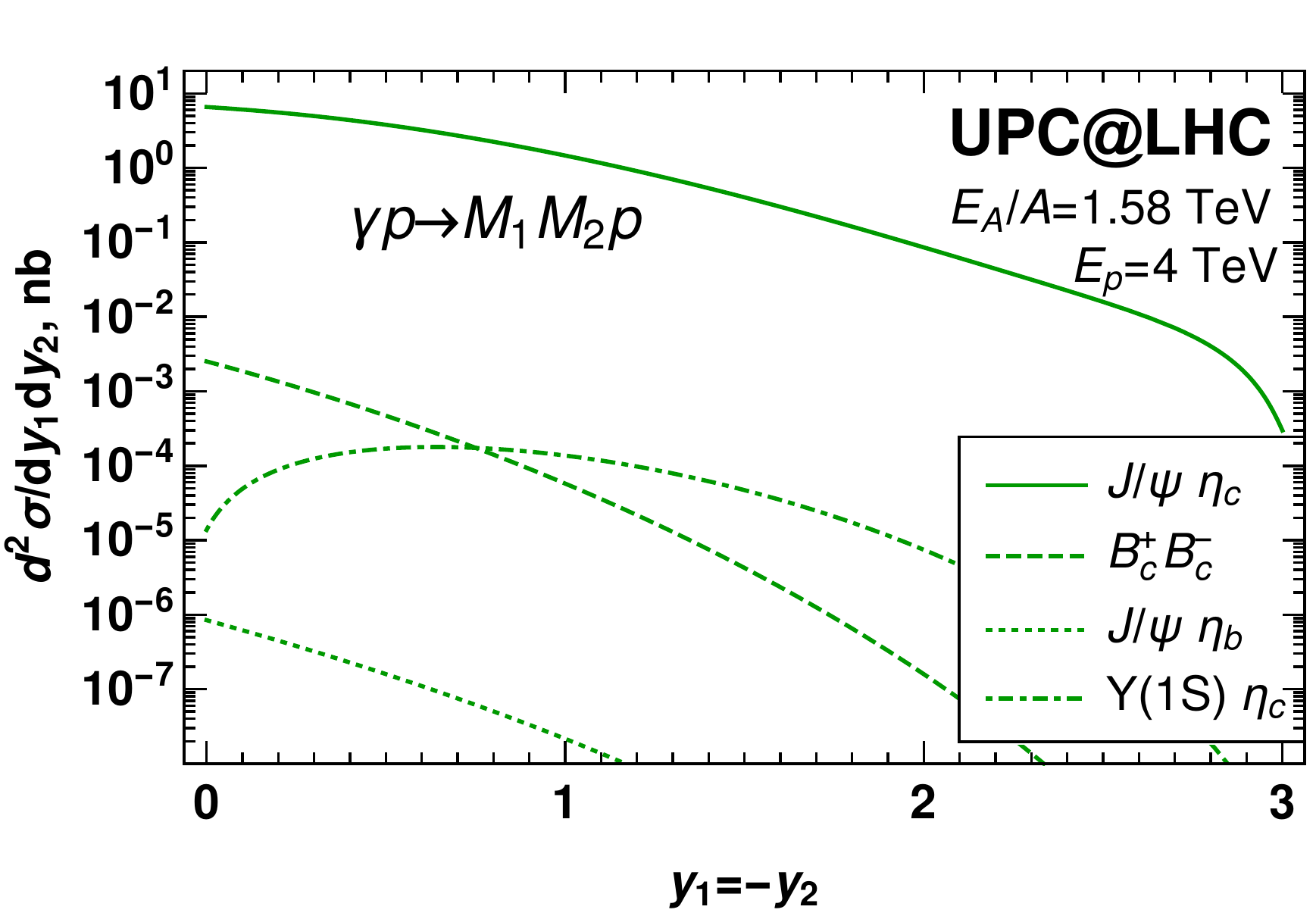}\includegraphics[width=9cm]{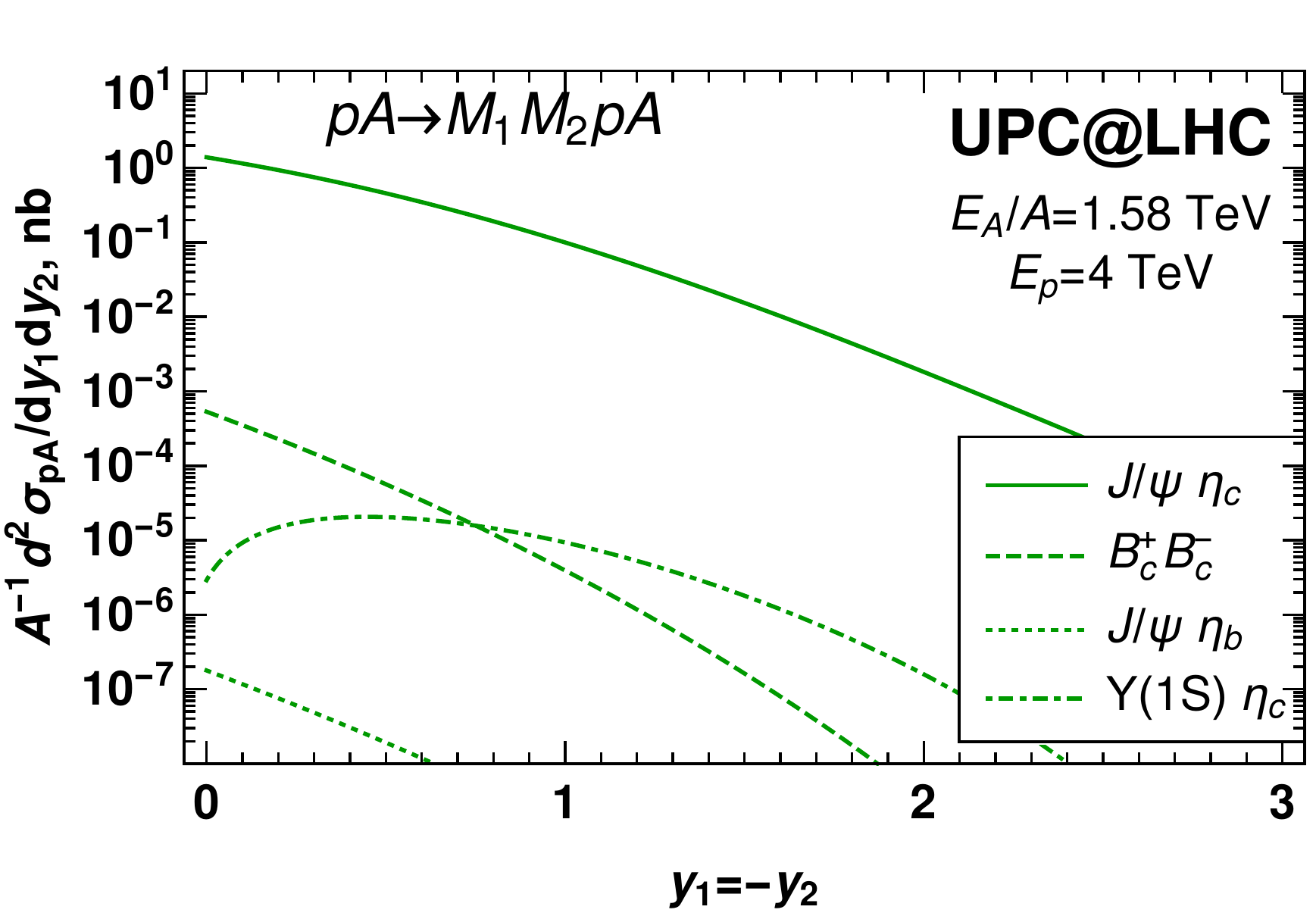}

\caption{\label{fig:Rapidities-UPC}The rapidity dependence of the photoproduction
cross-section in the kinematics of the ultraperipheral collisions at LHC.
The plots in the upper row correspond to a configuration with equal rapidities
of the produced quarkonia, $y_{1}=y_{2}$, whereas the lower row corresponds
to rapidities which differ by a sign in lab frame, $y_{1}=-y_{2}$.
In both rows the left plot corresponds to the \emph{photo}production
cross-section, and the right column shows predictions for the cross-section
of the full process $pA\to pAM_{1}M_{2}$, as defined in~(\ref{eq:EPA_q-2-1}).}
\end{figure}

Finally, in the Figure~\ref{fig:Rapidities} we show the dependence of the
cross-section on the quarkonia rapidities, integrated over the
transverse momenta of both heavy mesons. In the left panel we consider
the special case when both quarkonia are produced with the same rapidities
$y_{1}=y_{2}$ in the lab frame. The dependence on $y_{i}$ in this
setup merely reflects the dependence on the invariant photon-proton energy,
as could be seen from~(\ref{eq:W2}). In the right panel of the same
Figure~~\ref{fig:Rapidities} we show the dependence of the
cross-section on the rapidity difference $\Delta y$ between the two heavy
mesons. For the sake of definiteness we consider that both quarkonia
have opposite rapidities in the lab frame, $y_{1}=-y_{2}=\Delta y/2$.
In this setup the variable $\Delta y$ might be unambiguously related
to the invariant mass of the heavy quarkonia pair using~(\ref{eq:M12}).
We may observe that in this case the cross-section becomes suppressed
as a function of $\Delta y$, which illustrates the fact that the quarkonia
are predominantly produced with the same rapidities. Finally, in
Figure~\ref{fig:Rapidities-UPC} we show predictions for the pair
production in the kinematics of ultraperipheral collisions at LHC.
For the sake of definiteness, we consider proton-lead collisions.
Qualitatively the behavior of the cross-section is similar to that
of $ep$ production.

\section{Conclusions}

\label{sec:Conclusions}In this manuscript we have studied in detail the\emph{
}exclusive photoproduction of heavy quarkonia pairs, which include
bottom mesons. We focused on the leading order contribution, which
leads to production of charmonia and bottomonia pairs with opposite
$C$-parities. For $B_{c}^{+}B_{c}^{-}$ pairs, the $C$-parity does
not impose any constraints on the internal quantum numbers of quarkonia,
so the suggested mechanism might be used as a clean channel for studies
of (so far undiscovered) $B_{c}$ mesons with different internal quantum
numbers. The analysis of mixed charm-bottom pairs (e.g. $B_{c}^{+}B_{c}^{-}$
or $J/\psi\,\eta_{b}$, $\Upsilon\eta_{c}$ pairs) allows to single
out contributions of two main classes of diagrams
in the suggested production mechanism. In all cases the quarkonia
are produced with relatively small opposite transverse momenta $p_{T}$,
and small separation in rapidity: the kinematic which minimizes the
momentum transfer to the recoil proton and the invariant mass of the produced
pair. The dependence of the cross-section on azimuthal angle between
transverse momenta of produced quarkonia might present special interest,
since it allows to test the dependence of the dipole amplitude $N(x,\boldsymbol{r},\boldsymbol{b})$
on the relative angle between the dipole separation $\boldsymbol{r}$
and impact parameter $\boldsymbol{b}$. We estimated numerically the
cross-sections in the kinematics of ultraperipheral collisions at
LHC and the kinematics of the forthcoming Electron-Ion Collider. We
found that $J/\psi\,\eta_{c}$ and $B_{c}^{+}B_{c}^{-}$ might be
studied with reasonable precision in forthcoming experiments. The
production cross-sections of other quarkonia pairs, especially from
the all-bottom states (like e.g. $\Upsilon\eta_{b}$) are numerically
significantly smaller due to extra suppression by the heavy mass and a different
production mechanism. 

\section*{Acknowldgements}

We thank our colleagues at UTFSM university for encouraging discussions.
This research was partially supported by Proyecto ANID PIA/APOYO AFB180002
(Chile) and Fondecyt (Chile) grant 1220242. The research
of S. Andrade was also partially supported by the Fellowship Program
ANID BECAS/MAGÍSTER NACIONAL (Chile) 22200123. \textquotedbl Powered@NLHPC:
This research was partially supported by the supercomputing infrastructure
of the NLHPC (ECM-02)\textquotedbl .

\appendix

\section{Evaluation of the photon wave function}

\label{sec:WF}The evaluation of the photon wave function follows
the standard light--cone rules formulated in~\cite{Lepage:1980fj,Brodsky:1997de}.
The result for the $\bar{Q}Q$ component is well-known in the literature~\cite{Bjorken:1970ah,Dosch:1996ss}.
The wave function of the $\bar{Q}Q\bar{Q}Q$-component might be expressed
in terms of the wave function of $\bar{Q}Q$-component. The dominant
contribution~\footnote{We assume that there is no special cuts which select or enhance contributions
of events with large virtuality $Q^{2}$.} to the electroproduction and photoproduction in ultraperipheral kinematics
comes from quasireal transversely polarized photons, for this reason
in what follows we'll focus on the contribution of on-shell photons. The
expression for the momentum of the photon~(\ref{eq:qPhoton}) simplifies
in this case and has only light-cone components in the direction of plus-axis,
\begin{equation}
q\approx\left(q^{+},\,0,\,\boldsymbol{0}_{\perp}\right).
\end{equation}
The polarization vector of the transversely polarized photon is given
by
\begin{align}
\varepsilon_{T}^{\mu}(q) & \equiv\left(0,\,\frac{\boldsymbol{q}_{\perp}\cdot\boldsymbol{\boldsymbol{\varepsilon}}_{\gamma}}{q^{+}},\,\boldsymbol{\boldsymbol{\varepsilon}}_{\gamma}\right)\approx\left(0,\,0,\,\boldsymbol{\boldsymbol{\varepsilon}}_{\gamma}\right),\label{eq:eDef}\\
\boldsymbol{\boldsymbol{\varepsilon}}_{\gamma} & =\frac{1}{\sqrt{2}}\left(\begin{array}{c}
1\\
\pm i
\end{array}\right),\quad\gamma=\pm1.
\end{align}
where in~(\ref{eq:eDef}) we took into account that $\boldsymbol{q}_{\perp}=0$.
Before interaction with the target, the photon might fluctuate into a
virtual quark-antiquark pairs, as well as into gluons. In configuration
space the wave function of the $\bar{Q}Q$ state is given by~\cite{Bjorken:1970ah,Dosch:1996ss}
\begin{equation}
\Psi_{h\bar{h}}^{\lambda}\left(z,\,\boldsymbol{r}_{12},\,m_{q},a\right)=-\frac{2}{(2\pi)}\left[\left(z\delta_{\lambda,h}-(1-z)\delta_{\lambda,-h}\right)\delta_{h,-\bar{h}}i\boldsymbol{\varepsilon}_{\lambda}\cdot\nabla-\frac{m_{q}}{\sqrt{2}}\,{\rm sign}(h)\delta_{\lambda,h}\delta_{h,\bar{h}}\right]K_{0}\left(a\,\boldsymbol{r}_{12}\right).\label{eq:PsiSplitting}
\end{equation}
where $z$ is the fraction of the photon momentum carried by the quark,
and $\boldsymbol{r}_{12}$ is the transverse distance between quark
and antiquark. 

\begin{figure}
\includegraphics[scale=0.65]{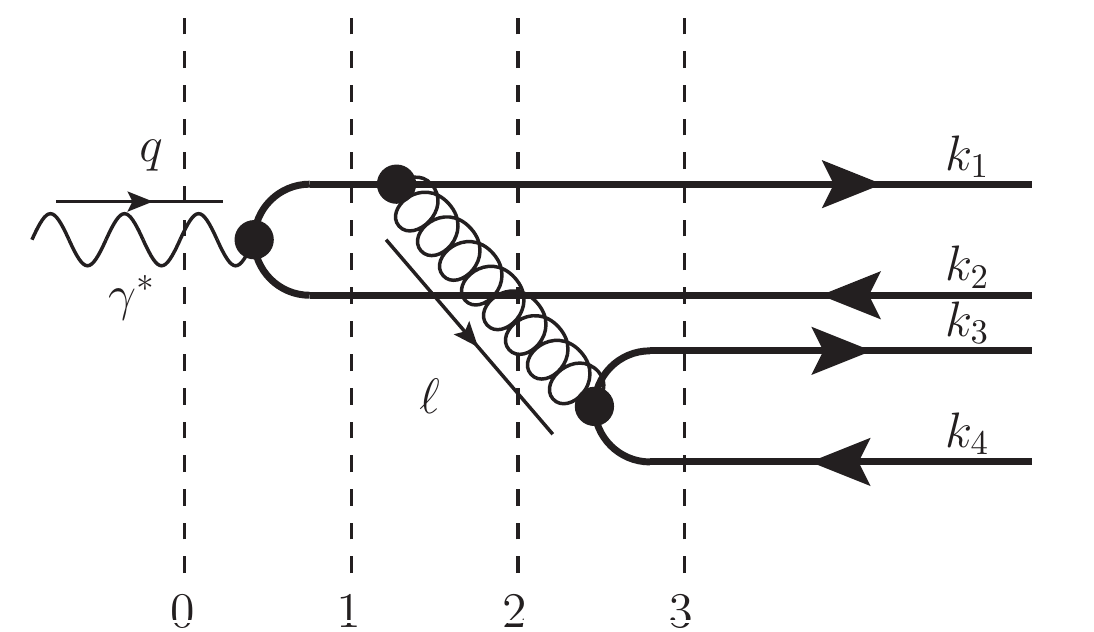}\includegraphics[scale=0.65]{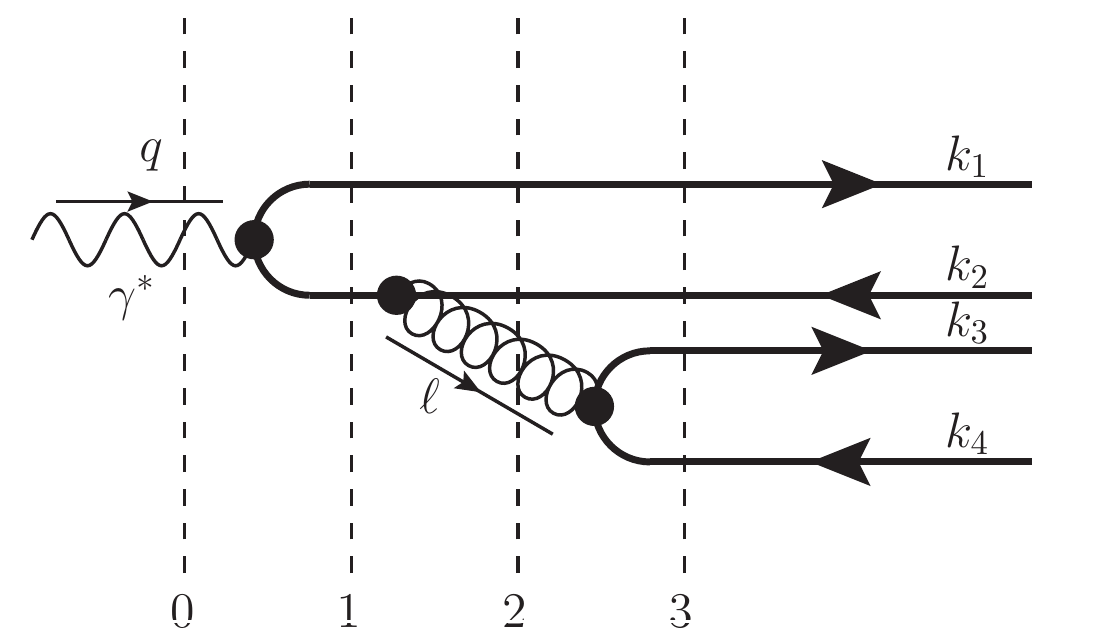}

\caption{\label{fig:Photoproduction-QQQQ}The leading order contribution to
the wave function $\psi_{\bar{Q}Q\bar{Q}Q}^{(\gamma)}$ defined in
the text. The momenta $k_{i}$ shown in the right-hand side are Fourier
conjugates of the coordinates $x_{i}$. It is implied that both diagrams
should be supplemented by all possible permutations of final state
quarks (see the text for more details).}
\end{figure}

The $\bar{Q}Q\bar{Q}Q$ component of the photon wave function has
been evaluated in detail in our earlier paper~\cite{Andrade:2022rbn},
and the final result will be given here for the sake of completeness.
In leading order over $\alpha_{s}$ the wave functions obtain
contributions from the two diagrams shown in the Figure~(\ref{fig:Photoproduction-QQQQ}).
For the sake of generality we will assume that the produced quark-antiquark
pairs have different flavors, and will use the notations $m_{1}$ for
the current mass of the quark line connected to a photon, and $m_{2}$
for the current masses of the quark-antiquark pair produced from the
virtual gluon. The evaluation of the diagrams follows the standard
rules of the light-cone perturbation theory~\cite{Lepage:1980fj,Brodsky:1997de}.
The wave function might be represented as a sum
\begin{equation}
\psi_{\bar{Q}Q\bar{Q}Q}^{(\gamma)}=\psi_{\bar{Q}Q\bar{Q}Q}^{(\gamma,{\rm noninst})}+\psi_{\bar{Q}Q\bar{Q}Q}^{(\gamma,{\rm inst})}
\end{equation}
where the first and the second terms correspond to contributions of
non-instantaneous and instantaneous parts of propagators of all virtual
particles, and for the sake of brevity we omitted color and helicity
indices of heavy quarks ($c_{i}$ and $a_{i}$ respectively). The
non-instantaneous contribution is given by the sum 
\begin{equation}
\psi_{\bar{Q}Q\bar{Q}Q}^{(\gamma,{\rm noninst})}\left(\left\{ \alpha_{i},\,\boldsymbol{x}_{i}\right\} \right)=A\left(\left\{ \alpha_{i},\,\boldsymbol{x}_{i}\right\} \right)+B\left(\left\{ \alpha_{i},\,\boldsymbol{x}_{i}\right\} \right).\label{eq:PsiFull}
\end{equation}

where

\begin{align}
A\left(\left\{ \alpha_{i},\,\boldsymbol{r}_{i}\right\} \right) & =-\frac{2e_{q}\alpha_{s}\left(\mu\right)\,\left(t_{a}\right)_{c_{1}c_{2}}\otimes\left(t_{a}\right)_{c_{3}c_{4}}}{\pi^{3}\left(1-\alpha_{1}-\alpha_{2}\right)^{2}\sqrt{\alpha_{1}\alpha_{2}}\,}\int\frac{q_{1}dq_{1}\,k_{2}dk_{2}}{\frac{\bar{\alpha}_{2}q_{1}^{2}}{\alpha_{1}\left(1-\alpha_{1}-\alpha_{2}\right)}+\frac{m_{1}^{2}\left(\alpha_{1}+\alpha_{2}\right)}{\alpha_{1}\alpha_{2}}+\frac{k_{2}^{2}}{\alpha_{2}\bar{\alpha}_{2}}}\times\label{eq:Amp_p_explicit-4-4}\\
 & \times\frac{1}{k_{2}^{2}+m_{1}^{2}}\sqrt{\frac{\alpha_{2}}{\alpha_{1}}}\left[\left(\alpha_{2}\delta_{\gamma,a_{2}}-\bar{\alpha}_{2}\delta_{\gamma,-a_{2}}\right)\left(\bar{\alpha}_{2}\delta_{\lambda,\,a_{1}}+\alpha_{1}\delta_{\lambda,-a_{1}}\right)\delta_{a_{1},-a_{2}}\times\right.\nonumber \\
 & \times\left(\boldsymbol{n}_{2,134}\cdot\boldsymbol{\varepsilon}_{\gamma}\right)\left(\boldsymbol{n}_{1,34}\cdot\boldsymbol{\varepsilon}_{\lambda}^{*}\right)k_{2}\,J_{1}\left(k_{2}\left|\boldsymbol{x}_{2}-\boldsymbol{b}_{134}\right|\right)q_{1}J_{1}\left(q_{1}\left|\boldsymbol{x}_{1}-\boldsymbol{b}_{34}\right|\right)+\nonumber \\
 & +\frac{m_{q}^{2}}{2}\,\delta_{\lambda,-a_{1}}\delta_{\gamma,a_{2}}\delta_{a_{1},-a_{2}}J_{0}\left(k_{2}\left|\boldsymbol{x}_{2}-\boldsymbol{b}_{134}\right|\right)J_{0}\left(q_{1}\left|\boldsymbol{x}_{1}-\boldsymbol{b}_{34}\right|\right)\frac{\left(1-\alpha_{1}-\alpha_{2}\right)^{2}}{1-\alpha_{2}}\nonumber \\
 & -\frac{im_{q}}{\sqrt{2}}\,{\rm sign}\left(a_{2}\right)\delta_{\gamma,a_{2}}\delta_{a_{1},a_{2}}\left(\bar{\alpha}_{2}\delta_{\lambda,\,a_{1}}+\alpha_{1}\delta_{\lambda,-a_{1}}\right)\times\nonumber \\
 & \times\boldsymbol{n}_{1,34}\cdot\boldsymbol{\varepsilon}_{\lambda}^{*}q_{1}J_{1}\left(q_{1}\left|\boldsymbol{x}_{1}-\boldsymbol{b}_{34}\right|\right)J_{0}\left(k_{2}\left|\boldsymbol{x}_{2}-\boldsymbol{b}_{134}\right|\right)\nonumber \\
 & -\frac{im_{q}}{\sqrt{2}}{\rm sign}\left(a_{1}\right)\delta_{\lambda,-a_{1}}\left(\alpha_{2}\delta_{\gamma,a_{2}}-\bar{\alpha}_{2}\delta_{\gamma,-a_{2}}\right)\delta_{a_{1},a_{2}}\frac{\left(1-\alpha_{1}-\alpha_{2}\right)^{2}}{1-\alpha_{2}}\times\nonumber \\
 & \times\left.\left(\boldsymbol{n}_{2,134}\cdot\boldsymbol{\varepsilon}_{\gamma}\right)k_{2}\,J_{1}\left(k_{2}\left|\boldsymbol{x}_{2}-\boldsymbol{b}_{134}\right|\right)J_{0}\left(q_{1}\left|\boldsymbol{x}_{1}-\boldsymbol{b}_{34}\right|\right)\right]\times\nonumber \\
 & \times\Psi_{a_{3},a_{4}}^{-\lambda}\left(\frac{\alpha_{3}}{\alpha_{3}+\alpha_{4}},\,\boldsymbol{r}_{34},\,m_{2},\,\sqrt{m_{2}^{2}+\frac{\alpha_{3}\alpha_{4}}{\alpha_{3}+\alpha_{4}}\left[\frac{\bar{\alpha}_{2}q_{1}^{2}}{\alpha_{1}\left(1-\alpha_{1}-\alpha_{2}\right)}+\frac{m_{1}^{2}\left(\alpha_{1}+\alpha_{2}\right)}{\alpha_{1}\alpha_{2}}+\frac{k_{2}^{2}}{\alpha_{2}\bar{\alpha}_{2}}\right]}\right)\nonumber 
\end{align}
and 
\[
B\left(\alpha_{1},\,\boldsymbol{x}_{1},\,\alpha_{2},\,\boldsymbol{x}_{2},\,\alpha_{3},\,\boldsymbol{x}_{3},\,\alpha_{4},\,\boldsymbol{x}_{4}\right)=-A\left(\alpha_{2},\,\boldsymbol{x}_{2},\,\alpha_{1},\,\boldsymbol{x}_{1},\,\alpha_{4},\,\boldsymbol{x}_{4},\,\alpha_{3},\,\boldsymbol{x}_{3}\right).
\]
The variable $\boldsymbol{b}_{j_{1}...j_{n}}$ corresponds to the center of mass position
of the $n$ partons $j_{1},\,...j_{n}$ . The variable
$\boldsymbol{n}_{i,j_{1}...j_{n}}=\left(\boldsymbol{x}_{i}-\boldsymbol{b}_{j_{1}...j_{n}}\right)/\left|\boldsymbol{x}_{i}-\boldsymbol{b}_{j_{1}...j_{n}}\right|$
is a unit vector pointing from quark $i$ towards the center-of-mass
of the system of quarks $j_{1}...j_{n}$. The tree-like structure of
the leading order diagrams 1, 2 in Fig.~\ref{fig:Photoproduction-QQQQ}
and iterative evaluation of the coordinate of the center of mass of
two partons $\boldsymbol{b}_{ij}=\left(\alpha_{i}\boldsymbol{r}_{i}+\alpha_{j}\boldsymbol{r}_{j}\right)/\left(\alpha_{i}+\alpha_{j}\right)$,
allows to reconstruct the transverse coordinates of all intermediate
partons. The variables $\boldsymbol{r}_{1}-\boldsymbol{b}_{34}$ and
$\boldsymbol{r}_{2}-\boldsymbol{b}_{34}$ physically have the meaning
of the relative distance between the recoil quark or antiquark and
the emitted gluon. Similarly, the variables $\boldsymbol{r}_{1}-\boldsymbol{b}_{234}$
and $\boldsymbol{r}_{2}-\boldsymbol{b}_{134}$ might be interpreted
as the size of $\bar{Q}Q$ pair produced right after splitting of the
incident photon. 

Similarly, for the instantaneous contributions it is possible to get
\begin{align}
\psi_{\bar{Q}Q\bar{Q}Q}^{(\gamma,{\rm inst.})}\left(\left\{ \alpha_{i},\,\boldsymbol{r}_{i}\right\} \right) & =A_{g}\left(\left\{ \alpha_{i},\,\boldsymbol{r}_{i}\right\} \right)+B_{g}\left(\left\{ \alpha_{i},\,\boldsymbol{r}_{i}\right\} \right)+\label{eq:PsiFull-1}\\
 & +A_{q}\left(\left\{ \alpha_{i},\,\boldsymbol{r}_{i}\right\} \right)+B_{q}\left(\left\{ \alpha_{i},\,\boldsymbol{r}_{i}\right\} \right),\nonumber 
\end{align}
where the subscript indices $q,g$ in the right-hand side denote the
parton propagator which should be taken instantaneous ($q$ for quark,
$g$ for gluon), and

\begin{align}
A_{g}\left(\left\{ \alpha_{i},\,\boldsymbol{r}_{i}\right\} \right) & =-\frac{e_{q}\alpha_{s}(m_{Q})\,\left(t_{a}\right)_{c_{1}c_{2}}\otimes\left(t_{a}\right)_{c_{3}c_{4}}}{\pi^{4}\left(1-\alpha_{1}-\alpha_{2}\right)^{3}\,}\int q_{1}dq_{1}\,k_{2}dk_{2}J_{0}\left(q_{1}\left|\boldsymbol{r}_{1}-\boldsymbol{b}_{34}\right|\right)\times\\
 & \times\frac{1}{\boldsymbol{k}_{2\perp}^{2}+m_{1}^{2}}\left[\left(\alpha_{2}\delta_{\gamma,a_{1}}-\bar{\alpha}_{2}\delta_{a_{1},-\gamma}\right)\delta_{a_{1},-a_{2}}i\boldsymbol{n}_{2,134}\cdot\boldsymbol{\varepsilon}_{\gamma}k_{2}J_{1}\left(k_{2}\left|\boldsymbol{r}_{2}-\boldsymbol{b}_{134}\right|\right)\right.\nonumber \\
 & \left.+\frac{m_{q}}{\sqrt{2}}\,{\rm sign}\left(a_{1}\right)\delta_{\gamma,a_{1}}\delta_{a_{1},a_{2}}J_{0}\left(k_{2}\left|\boldsymbol{r}_{2}-\boldsymbol{b}_{134}\right|\right)\right]\alpha_{3}\alpha_{4}\delta_{a_{3},-a_{4}}K_{0}\left(a_{34}\,\boldsymbol{r}_{34}\right).\nonumber 
\end{align}
\begin{align}
A_{q}\left(\left\{ \alpha_{i},\,\boldsymbol{r}_{i}\right\} \right) & =-\frac{e_{q}\alpha_{s}\left(m_{q}\right)\,\left(t_{a}\right)_{c_{1}c_{2}}\otimes\left(t_{a}\right)_{c_{3}c_{4}}}{2\pi^{4}\left(1-\alpha_{1}-\alpha_{2}\right)^{2}\bar{\alpha}_{2}}\delta_{a_{1},-a_{2}}\delta_{\gamma,-a_{1}}\int q_{1}dq_{1}\,k_{2}dk_{2}\frac{J_{0}\left(q_{1}\left|\boldsymbol{r}_{1}-\boldsymbol{b}_{34}\right|\right)J_{0}\left(k_{2}\left|\boldsymbol{r}_{2}-\boldsymbol{b}_{134}\right|\right)}{D_{2}\left(\alpha_{1},\boldsymbol{k}_{1};\alpha_{2},\,\boldsymbol{k}_{2}\right)}\times\\
 & \times\left[-\left(\alpha_{3}\delta_{-\gamma,a_{3}}-\alpha_{4}\delta_{\gamma,a_{3}}\right)\delta_{a_{3},-a_{4}}i\boldsymbol{\varepsilon}_{\gamma}\cdot\boldsymbol{n}_{34}a_{34}K_{1}\left(a_{34}\,\boldsymbol{r}_{34}\right)-\frac{m_{q}(\alpha_{3}+\alpha_{4})}{\sqrt{2}}\,{\rm sign}(a_{3})\delta_{\gamma,-a_{3}}\delta_{a_{3},a_{4}}K_{0}\left(a_{34}\,\boldsymbol{r}_{34}\right)\right]\nonumber \\
 & a_{34}\left(q_{1},\,k_{2}\right)\equiv\sqrt{m_{2}^{2}+\frac{\alpha_{3}\alpha_{4}}{\alpha_{3}+\alpha_{4}}\left[\frac{\bar{\alpha}_{2}q_{1}^{2}}{\alpha_{1}\left(1-\alpha_{1}-\alpha_{2}\right)}+\frac{m_{1}^{2}\left(\alpha_{1}+\alpha_{2}\right)}{\alpha_{1}\alpha_{2}}+\frac{k_{2}^{2}}{\alpha_{2}\bar{\alpha}_{2}}\right]}
\end{align}
and the functions $B_{q},\,B_{g}$ might be obtained from $A_{q},A_{g}$
using 
\begin{equation}
B_{i}\left(\alpha_{1},\,\boldsymbol{x}_{1},\,\alpha_{2},\,\boldsymbol{x}_{2},\,\alpha_{3},\,\boldsymbol{x}_{3},\,\alpha_{4},\,\boldsymbol{x}_{4}\right)=-A_{i}\left(\alpha_{2},\,\boldsymbol{x}_{2},\,\alpha_{1},\,\boldsymbol{x}_{1},\,\alpha_{4},\,\boldsymbol{x}_{4},\,\alpha_{3},\,\boldsymbol{x}_{3}\right),\quad i=q,g.
\end{equation}

 \end{document}